# Energy, linear momentum, and angular momentum exchange between an electromagnetic wave-packet and a small particle

Masud Mansuripur

College of Optical Sciences, The University of Arizona, Tucson

**Abstract.** Invoking Maxwell's classical electrodynamics in conjunction with expressions for the electromagnetic (EM) energy, momentum, force, and torque, we use a few simple examples to demonstrate the nature of linear and angular momentum exchange between a wave-packet and a small spherical particle. The linear and angular momenta of the EM field, when absorbed by the particle, will be seen to elicit different responses from the particle.

**1. Introduction**. The present paper aims to illustrate the mechanisms of exchange of energy as well as linear and angular momenta between an electromagnetic (EM) wave-packet propagating in free space and a small (spherical) particle that acquires an induced polarization upon encountering the wave-packet; see Fig.1. The packet will continue to propagate beyond the particle, with its energy and momenta intact. The particle, while excited, radiates an EM wave that carries energy and angular momentum—but no linear momentum. At the same time, the particle may absorb (permanently) energy as well as linear and angular momenta from the incident packet. If, in advance of the packet's arrival, the particle happens to be in an excited state (e.g., pumped gain medium), its stored internal energy will be released by stimulated emission into the surrounding environment. Many interesting phenomena associated with the radiation absorption and emission processes result from interference between the wave-packet and the EM field that has been radiated into the particle's surrounding space. We examine a few cases of practical interest.

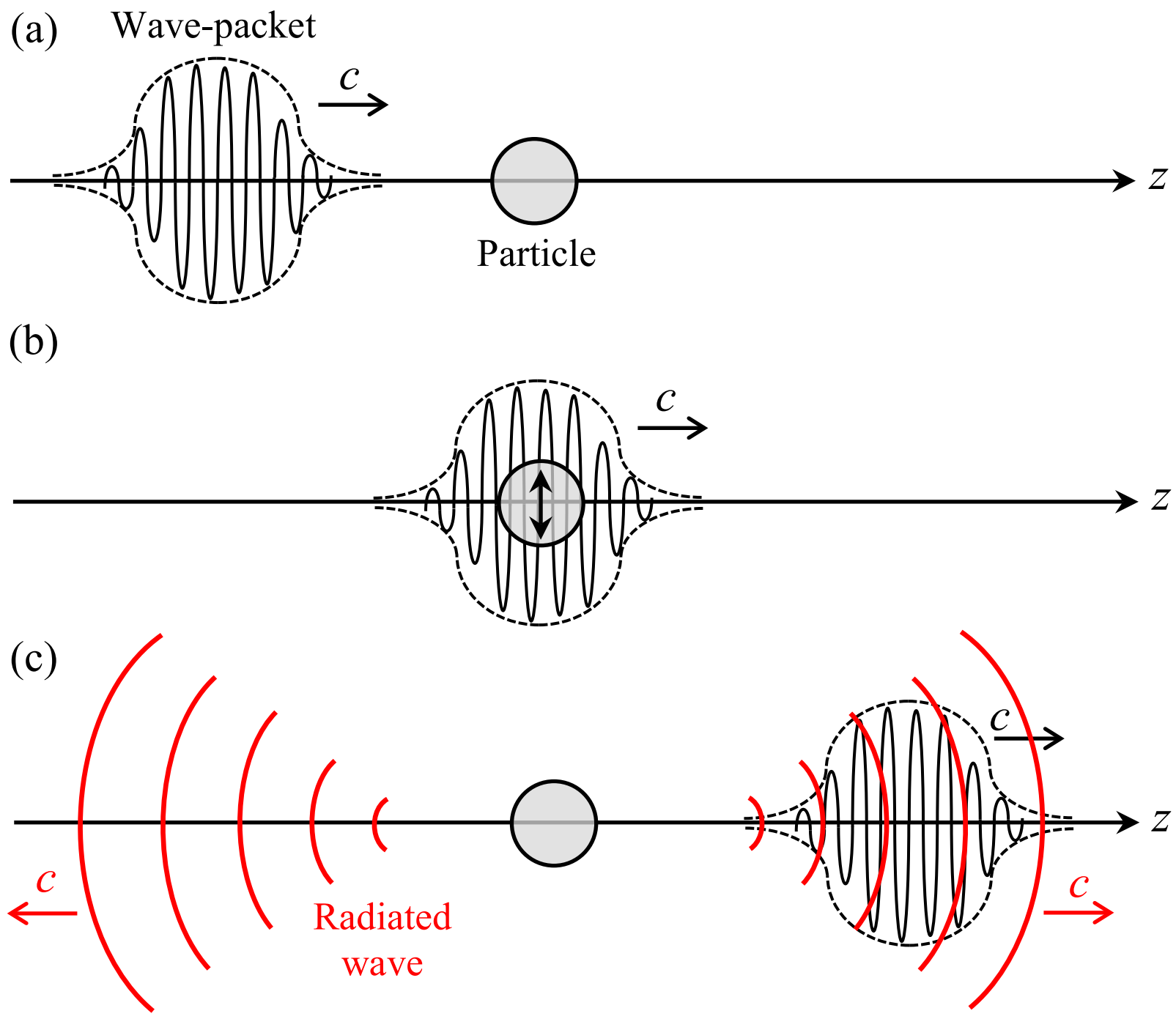

**Fig.1**. (a) A wave-packet propagating in free space arrives at a small spherical particle of radius $R$ and dielectric susceptibility $\varepsilon_0\chi(\omega)$. (b) The particle responds to the wave-packet's $E$-field, acquiring an induced polarization $\boldsymbol{P}(t)$ for the duration of the packet. (c) The wave-packet continues on its journey, unhindered, while the dipolar radiation expands in the space surrounding the particle. Interference between the wave-packet and the (expanding) radiated field accounts for a substantial contribution to the EM energy, linear momentum, and angular momentum of the overall EM field.

The goal of the paper is not so much to reveal new physics as it is to shed light on the known physics of interaction between a small particle and an EM wave-packet. By separating the contributions of the incident wave from those of the field that is radiated by the excited particle, we hope to clarify the mechanisms of exchange of energy and momentum between EM fields and material media. To simplify the argument, the particle will be assumed to be spherical and much smaller than the wavelength of the incident light, but of course the argument is quite general and can, in principle, be applied to more complex waveforms and extended material objects.

The standard classical theory of electrodynamics forms the basis for all the calculations reported in the present paper. The fundamental assumptions, the methodology, and the final results of our analysis are described in the sections that follow. The algebraic details of the calculations, however, are too tedious to include in the main body of the paper without distracting from the main results; these details are relegated to several appendices that appear at the end of the paper.

**2. Formalism**. The electric and magnetic fields of an infinitely-wide, uniformly-polarized, finite duration wave-packet propagating in free space along the $z$-axis may be written as a superposition of plane-waves having $k$-vectors $(k_x, k_y, k_z) = (0, 0, \omega/c)$ and $E$-field amplitudes $\vec{\mathbb{E}}(\omega) = \mathbb{E}_x(\omega)\hat{\boldsymbol{x}} + \mathbb{E}_y(\omega)\hat{\boldsymbol{y}}$, namely,

$$\boldsymbol{E}(\boldsymbol{r}, t) = \frac{1}{2\pi}\int_{-\infty}^{\infty} \vec{\mathbb{E}}(\omega)\exp[\mathrm{i}(\omega/c)(z - ct)]\,\mathrm{d}\omega, \tag{1a}$$

$$\boldsymbol{H}(\boldsymbol{r}, t) = \frac{1}{2\pi Z_0}\int_{-\infty}^{\infty} \hat{\boldsymbol{z}} \times \vec{\mathbb{E}}(\omega)\exp[\mathrm{i}(\omega/c)(z - ct)]\,\mathrm{d}\omega. \tag{1b}$$

In the above equations, $c = 1/\sqrt{\mu_0\varepsilon_0}$ is the speed of light in vacuum, and $Z_0 = \sqrt{\mu_0/\varepsilon_0}$ is the impedance of free space [1,2]. Appendix A shows that the energy and linear momentum content of the packet (per unit cross-sectional area) are given by

$$\mathcal{E} = \tfrac{1}{2}\varepsilon_0\int_{-\infty}^{\infty}\boldsymbol{E}(\boldsymbol{r}, t)\cdot\boldsymbol{E}(\boldsymbol{r}, t)\mathrm{d}z + \tfrac{1}{2}\mu_0\int_{-\infty}^{\infty}\boldsymbol{H}(\boldsymbol{r}, t)\cdot\boldsymbol{H}(\boldsymbol{r}, t)\mathrm{d}z = \frac{1}{2\pi Z_0}\int_{-\infty}^{\infty}\vec{\mathbb{E}}(\omega)\cdot\vec{\mathbb{E}}^*(\omega)\mathrm{d}\omega. \tag{2}$$

$$\boldsymbol{p} = \int_{-\infty}^{\infty}[\boldsymbol{E}(\boldsymbol{r}, t)\times\boldsymbol{H}(\boldsymbol{r}, t)/c^2]\mathrm{d}z = (\varepsilon_0/2\pi)\hat{\boldsymbol{z}}\int_{-\infty}^{\infty}\vec{\mathbb{E}}(\omega)\cdot\vec{\mathbb{E}}^*(\omega)\mathrm{d}\omega = (\mathcal{E}/c)\hat{\boldsymbol{z}}. \tag{3}$$

Computing the angular momentum [3-6] of the wave-packet requires a modified field profile, as the uniform cross-section of the infinitely-wide packet defined by Eq.(1) is not suited for angular momentum calculations [7,8]. For this we must use the following finite-duration *and* finite cross-section wave-packet:

$$\boldsymbol{E}(\boldsymbol{r}, t) = \frac{1}{(2\pi)^3}\iiint_{-\infty}^{\infty}\vec{\mathbb{E}}(\boldsymbol{k})\exp[\mathrm{i}(\boldsymbol{k}\cdot\boldsymbol{r} - \omega t)]\,\mathrm{d}k_x\mathrm{d}k_y\mathrm{d}k_z, \tag{4a}$$

$$\boldsymbol{H}(\boldsymbol{r}, t) = \frac{1}{(2\pi)^3 Z_0}\iiint_{-\infty}^{\infty}(c\boldsymbol{k}/\omega)\times\vec{\mathbb{E}}(\boldsymbol{k})\exp[\mathrm{i}(\boldsymbol{k}\cdot\boldsymbol{r} - \omega t)]\,\mathrm{d}k_x\mathrm{d}k_y\mathrm{d}k_z. \tag{4b}$$

The frequency $\omega$ appearing in Eq.(4) is no longer an independent variable, but tied to the $k$-vector as

$$\omega = ck_z\sqrt{1 + (k_x/k_z)^2 + (k_y/k_z)^2}. \tag{4c}$$

As shown in Appendix B, the angular momentum of the wave-packet defined by Eq.(4) is given in its entirety by

$$\boldsymbol{\mathcal{L}} = \int \boldsymbol{r}\times[\boldsymbol{E}(\boldsymbol{r}, t)\times\boldsymbol{H}(\boldsymbol{r}, t)/c^2]\mathrm{d}x\mathrm{d}y\mathrm{d}z$$
$$= \frac{\mathrm{i}\varepsilon_0}{(2\pi)^3}\iiint_{-\infty}^{\infty}\omega^{-1}\{[(\vec{\mathbb{E}}\cdot\partial_{k_x}\vec{\mathbb{E}}^*)\hat{\boldsymbol{x}} + (\vec{\mathbb{E}}\cdot\partial_{k_y}\vec{\mathbb{E}}^*)\hat{\boldsymbol{y}} + (\vec{\mathbb{E}}\cdot\partial_{k_z}\vec{\mathbb{E}}^*)\hat{\boldsymbol{z}}]\times\boldsymbol{k} + (\vec{\mathbb{E}}^*\times\vec{\mathbb{E}})\}\mathrm{d}k_x\mathrm{d}k_y\mathrm{d}k_z. \tag{5}$$

In the above equation, the second term of the integrand, namely, $\omega^{-1}\vec{\mathbb{E}}^*\times\vec{\mathbb{E}}$, corresponds to *spin*, while the remaining term represents the *orbital* and *extrinsic* angular momenta of the wave-packet [7-11].

Equation (4) may also be used to find the total energy and total linear momentum of an arbitrary wave-packet (see Appendix B), thus generalizing the results given previously (for an infinitely-wide packet of uniform cross-section) in Eqs.(2) and (3), as follows:

$$\mathcal{E} = \frac{\varepsilon_0}{(2\pi)^3}\iiint_{-\infty}^{\infty}\vec{\mathbb{E}}(\boldsymbol{k})\cdot\vec{\mathbb{E}}^*(\boldsymbol{k})\mathrm{d}k_x\mathrm{d}k_y\mathrm{d}k_z. \tag{6}$$

$$\boldsymbol{p} = \frac{\varepsilon_0}{(2\pi)^3}\iiint_{-\infty}^{\infty}\omega^{-1}[\vec{\mathbb{E}}(\boldsymbol{k})\cdot\vec{\mathbb{E}}^*(\boldsymbol{k})]\boldsymbol{k}\mathrm{d}k_x\mathrm{d}k_y\mathrm{d}k_z. \tag{7}$$

Calculations pertaining to the dipolar field radiated by the particle – as well as those arising from interference between the radiated field and the undisturbed wave-packet – are substantially more complicated; results of these calculations (along with detailed analyses) will be reported in the following sections. In the remainder of the present section, we limit our discussion to a procedural issue in the context of a general description of the problem under investigation. As a procedural matter, note that, if the wave-packet's $E$ and $H$ fields are denoted by the superscript (wp), and those of the dipole radiation by the superscript (dr), then the total linear momentum of the EM field at $t = t_0$ may be expressed as

$$\boldsymbol{p}^{(\text{total})}(t_0) = \frac{1}{c^2}\iiint_{-\infty}^{\infty}\left[\boldsymbol{E}^{(\text{wp})}(\boldsymbol{r},t_0) + \boldsymbol{E}^{(\text{dr})}(\boldsymbol{r},t_0)\right] \times \left[\boldsymbol{H}^{(\text{wp})}(\boldsymbol{r},t_0) + \boldsymbol{H}^{(\text{dr})}(\boldsymbol{r},t_0)\right]\mathrm{d}x\mathrm{d}y\mathrm{d}z$$

$$= \boldsymbol{p}^{(\text{wp})}(t_0) + \boldsymbol{p}^{(\text{dr})}(t_0) + \frac{1}{c^2}\iiint_{-\infty}^{\infty}\left[\boldsymbol{E}^{(\text{wp})} \times \boldsymbol{H}^{(\text{dr})} + \boldsymbol{E}^{(\text{dr})} \times \boldsymbol{H}^{(\text{wp})}\right]\mathrm{d}x\mathrm{d}y\mathrm{d}z. \qquad (8)$$

The latter integral in Eq.(8) represents the contribution to total EM momentum of the cross-terms produced by interference between the wave-packet and the dipole radiation. Similar expressions can be constructed for the contributions of cross-terms to total EM energy and angular momentum of the system.

We mention in passing that the dielectric susceptibility $\chi(\omega)$ of a small spherical particle of volume $v = (4\pi/3)R^3$, excited by an EM wave at the excitation frequency $\omega$ (corresponding to a vacuum wavelength of $\lambda = 2\pi c/\omega$), is given by the single-electron Lorentz oscillator model [1,2] having oscillator strength $f_{\text{os}}$, plasma frequency $\omega_p$, resonance frequency $\omega_0$, and damping coefficient $\gamma$, as follows [4]:

$$\chi(\omega) = \frac{f_{\text{os}}\omega_p^2}{\omega_0^2 - \omega^2 - \mathrm{i}\left[\gamma\omega + (4\pi^2 f_{\text{os}}\omega_p^2/3)(v/\lambda^3)\right]}. \qquad (9)$$

The term $(4\pi^2 f\omega_p^2/3)(v/\lambda^3)$ appearing in the denominator of $\chi$ is contributed by radiation resistance.

**3. Solving Maxwell's equations in the Fourier domain**. In general, the electric field of a plane-wave having frequency $\omega$ and $k$-vector $\boldsymbol{k}$, is written as $\boldsymbol{E}(\boldsymbol{r},t) = \vec{\mathbb{E}}(\boldsymbol{k},\omega)\exp[\mathrm{i}(\boldsymbol{k}\cdot\boldsymbol{r} - \omega t)]$. Here $\vec{\mathbb{E}}$, a complex vector function of $\boldsymbol{k}$ and $\omega$, is the plane-wave's $E$-field amplitude, which may be written in terms of its real and imaginary components as $\vec{\mathbb{E}} = \vec{\mathbb{E}}_{\text{R}} + \mathrm{i}\vec{\mathbb{E}}_{\text{I}}$. Considering that the $E$-field, being a physical entity, is always real, its Fourier transform must be Hermitian, that is, $\vec{\mathbb{E}}(-\boldsymbol{k},-\omega) = \vec{\mathbb{E}}^*(\boldsymbol{k},\omega)$. Similar expressions can be written for the magnetic fields $\boldsymbol{B}(\boldsymbol{r},t)$ and $\boldsymbol{H}(\boldsymbol{r},t)$, as well as for all the sources. In what follows, we shall assume that free charge-density $\rho_{\text{free}}(\boldsymbol{r},t)$, free current-density $\boldsymbol{J}_{\text{free}}(\boldsymbol{r},t)$, and magnetization $\boldsymbol{M}(\boldsymbol{r},t)$ are absent from the system and that, therefore, the only source term that needs to be taken into account is the polarization $\boldsymbol{P}(\boldsymbol{r},t)$. We will then have the displacement field $\boldsymbol{D} = \varepsilon_0\boldsymbol{E} + \boldsymbol{P}$ and the magnetic induction $\boldsymbol{B} = \mu_0\boldsymbol{H} + \boldsymbol{M} = \mu_0\boldsymbol{H}$. Maxwell's equations may thus be written in the space-time domain $(\boldsymbol{r},t)$, and also in the Fourier domain $(\boldsymbol{k},\omega)$, as follows:

$$\boldsymbol{\nabla}\cdot\boldsymbol{D}(\boldsymbol{r},t) = 0 \quad \rightarrow \quad \boldsymbol{k}\cdot[\varepsilon_0\vec{\mathbb{E}}(\boldsymbol{k},\omega) + \vec{\mathbb{P}}(\boldsymbol{k},\omega)] = 0. \qquad (10a)$$

$$\boldsymbol{\nabla}\times\boldsymbol{H}(\boldsymbol{r},t) = \partial_t\boldsymbol{D}(\boldsymbol{r},t) \quad \rightarrow \quad \boldsymbol{k}\times\vec{\mathbb{H}}(\boldsymbol{k},\omega) = -\omega[\varepsilon_0\vec{\mathbb{E}}(\boldsymbol{k},\omega) + \vec{\mathbb{P}}(\boldsymbol{k},\omega)]. \qquad (10b)$$

$$\boldsymbol{\nabla}\times\boldsymbol{E}(\boldsymbol{r},t) = -\partial_t\boldsymbol{B}(\boldsymbol{r},t) \quad \rightarrow \quad \boldsymbol{k}\times\vec{\mathbb{E}}(\boldsymbol{k},\omega) = \mu_0\omega\vec{\mathbb{H}}(\boldsymbol{k},\omega). \qquad (10c)$$

Therefore,

$$\boldsymbol{k}\times(\boldsymbol{k}\times\vec{\mathbb{E}}) = -\mu_0\omega^2(\varepsilon_0\vec{\mathbb{E}} + \vec{\mathbb{P}}) \quad \rightarrow \quad (\boldsymbol{k}\cdot\vec{\mathbb{E}})\boldsymbol{k} - k^2\vec{\mathbb{E}} = -(\omega/c)^2\left(\vec{\mathbb{E}} + \varepsilon_0^{-1}\vec{\mathbb{P}}\right)$$

$$\rightarrow \quad \vec{\mathbb{E}}(\boldsymbol{k},\omega) = \frac{(\omega/c)^2\vec{\mathbb{P}} - (\vec{\mathbb{P}}\cdot\boldsymbol{k})\boldsymbol{k}}{\varepsilon_0[k^2 - (\omega/c)^2]}. \qquad (11a)$$

$$\rightarrow \quad \vec{\mathbb{H}}(\boldsymbol{k},\omega) = \frac{\boldsymbol{k}\times\vec{\mathbb{E}}}{\mu_0\omega} = \frac{\boldsymbol{k}\times\omega\vec{\mathbb{P}}}{k^2 - (\omega/c)^2} \qquad (11b)$$

Note that there is one questionable step in the above derivation, which has to do with the fact that at $k = \pm\omega/c$, division by $k^2 - (\omega/c)^2$ may not be allowed. To better understand the problem, suppose that the radiation source $\boldsymbol{P}(\boldsymbol{r},t)$ is a uniformly polarized spherical particle of radius $R$ centered at the origin of coordinates, having the time-dependent polarization $\boldsymbol{P}(t)$, which vanishes outside the interval $0 \le t \le T$. The Fourier transform of $\boldsymbol{P}(\boldsymbol{r},t)$ is thus given by

$$\vec{\mathbb{P}}(\boldsymbol{k},\omega) = \int_{r=0}^{R}\int_{\theta=0}^{\pi} 2\pi r\sin\theta \exp(-\mathrm{i}\boldsymbol{k}\cdot\boldsymbol{r})\, r\mathrm{d}\theta\mathrm{d}r \int_{t=0}^{T} \boldsymbol{P}(t)\exp(\mathrm{i}\omega t)\,\mathrm{d}t$$

$$= 2\pi\vec{\mathbb{P}}(\omega)\int_0^R r^2\left[\int_0^\pi \sin\theta\exp(-\mathrm{i}kr\cos\theta)\,\mathrm{d}\theta\right]\mathrm{d}r = (4\pi/k)\vec{\mathbb{P}}(\omega)\int_0^R r\sin(kr)\,\mathrm{d}r$$

$$= 4\pi\vec{\mathbb{P}}(\omega)\,[\sin(kR) - kR\cos(kR)]/k^3. \tag{12}$$

Substitution into Eqs.(11) followed by an inverse Fourier transformation now yields

$$\boldsymbol{E}(\boldsymbol{r},t) = \frac{4\pi}{(2\pi)^4\varepsilon_0}\int_{-\infty}^{\infty} \frac{(\omega/c)^2\vec{\mathbb{P}} - (\vec{\mathbb{P}}\cdot\boldsymbol{k})\boldsymbol{k}}{k^2-(\omega/c)^2}\exp[\mathrm{i}(\boldsymbol{k}\cdot\boldsymbol{r}-\omega t)]\,\mathrm{d}\boldsymbol{k}\mathrm{d}\omega$$

$$= \frac{1}{4\pi^3\varepsilon_0}\int_{-\infty}^{\infty}\left[\frac{\sin(kR)-kR\cos(kR)}{k^3}\right]\exp(\mathrm{i}\boldsymbol{k}\cdot\boldsymbol{r})\left\{\int_{-\infty}^{\infty}\frac{(\omega/c)^2\vec{\mathbb{P}}(\omega) - [\vec{\mathbb{P}}(\omega)\cdot\boldsymbol{k}]\boldsymbol{k}}{k^2-(\omega/c)^2}\exp(-\mathrm{i}\omega t)\mathrm{d}\omega\right\}\mathrm{d}\boldsymbol{k}. \tag{13a}$$

$$\boldsymbol{H}(\boldsymbol{r},t) = \frac{1}{4\pi^3}\int_{-\infty}^{\infty}\left[\frac{\sin(kR)-kR\cos(kR)}{k^3}\right]\exp(\mathrm{i}\boldsymbol{k}\cdot\boldsymbol{r})\left\{\int_{-\infty}^{\infty}\frac{\boldsymbol{k}\times\omega\vec{\mathbb{P}}(\omega)}{k^2-(\omega/c)^2}\exp(-\mathrm{i}\omega t)\mathrm{d}\omega\right\}\mathrm{d}\boldsymbol{k}. \tag{13b}$$

It is clear from the above equations that $1/[k^2-(\omega/c)^2]$ acts in the $\omega$ domain as the transfer function of any linear filter would act on the spectral distribution of an input, so that, upon inverse Fourier transformation, a filtered version of the input, a function of $t$, is obtained. The filter must be causal because the response cannot begin before the excitation function $\boldsymbol{P}(\boldsymbol{r},t)$ has started. We conjecture, therefore, that the impulse-response of the filter must have the form $Q_k(t) = \mathrm{step}(t)\sin(ckt)$, whose Fourier transform is readily obtained as follows:

$$\mathcal{F}\{Q_k(t)\} = lim_{\alpha\to 0}\int_0^\infty \exp(-\alpha t)\sin(ckt)\exp(\mathrm{i}\omega t)\,\mathrm{d}t$$

$$= lim_{\alpha\to 0}\frac{1}{2\mathrm{i}}\left\{\int_0^\infty \exp\{-[\alpha-\mathrm{i}(ck+\omega)]t\}\,\mathrm{d}t - \int_0^\infty \exp\{-[\alpha+\mathrm{i}(ck-\omega)]t\}\,\mathrm{d}t\right\}$$

$$= lim_{\alpha\to 0}\frac{1}{2\mathrm{i}}\left\{\frac{1}{\alpha-\mathrm{i}(ck+\omega)} - \frac{1}{\alpha+\mathrm{i}(ck-\omega)}\right\} = lim_{\alpha\to 0}\frac{1}{2\mathrm{i}}\left\{\frac{\alpha+\mathrm{i}(\omega+ck)}{\alpha^2+(\omega+ck)^2} - \frac{\alpha+\mathrm{i}(\omega-ck)}{\alpha^2+(\omega-ck)^2}\right\}$$

$$= \frac{k/c}{k^2-(\omega/c)^2} + \tfrac{1}{2}\mathrm{i}\pi[\delta(\omega-ck) - \delta(\omega+ck)]\cdot \tag{14}$$

The pair of delta-functions in Eq.(14) is thus seen to be a necessary companions of the transfer function $1/[k^2-(\omega/c)^2]$ in Eq.(11), if the causality of the filter is to be assured. These $\delta$-functions are located precisely where the denominator on the right-hand side of Eqs.(11) vanishes, which is where one suspects that division by zero could be problematic. The addition of this pair of $\delta$-functions to the transfer function $1/[k^2-(\omega/c)^2]$ is all that is needed to ensure the accuracy of Eqs.(11), and, by extension, that of Eqs.(13).

Consider now a finite-duration function $F(t)$, where $F(t)\neq 0$ only during the interval $0\le t\le T$. Upon entering the aforementioned filter and getting convolved with its causal impulse-response $Q_k(t)$, we find that the resulting convolution will be zero for $t<0$. The Fourier transform of the convolution is given by

$$\mathcal{F}\{F(t)*Q_k(t)\} = \int_{-\infty}^{\infty}\left[\int_{-\infty}^{\infty}F(t')Q_k(t-t')\mathrm{d}t'\right]\exp(\mathrm{i}\omega t)\,\mathrm{d}t$$

$$= \int_{-\infty}^{\infty}F(t')\left[\int_{-\infty}^{\infty}Q_k(t-t')\exp(\mathrm{i}\omega t)\,\mathrm{d}t\right]\mathrm{d}t'$$

$$= \mathbb{Q}_k(\omega)\int_{-\infty}^{\infty}F(t')\exp(\mathrm{i}\omega t')\,\mathrm{d}t' = \mathbb{F}(\omega)\mathbb{Q}_k(\omega). \tag{15}$$

If we pick the observation time $t_0 > T$, and assume that $F(t)$ is real-valued—in which case $\mathbb{F}(\omega)$ will be Hermitian—we will have

$$F(t) * Q_k(t)|_{t=t_0} = \int_0^T F(t)\sin[ck(t_0 - t)]\,\mathrm{d}t$$

$$= \tfrac{1}{2\mathrm{i}}\exp(\mathrm{i}ckt_0)\int_{-\infty}^{\infty} F(t)\exp(-\mathrm{i}ckt)\,\mathrm{d}t - \tfrac{1}{2\mathrm{i}}\exp(-\mathrm{i}ckt_0)\int_{-\infty}^{\infty} F(t)\exp(\mathrm{i}ckt)\,\mathrm{d}t$$

$$= \tfrac{1}{2\mathrm{i}}\exp(\mathrm{i}ckt_0)\,\mathbb{F}^*(ck) - \tfrac{1}{2\mathrm{i}}\exp(-\mathrm{i}ckt_0)\,\mathbb{F}(ck) \quad \leftarrow \boxed{\mathbb{F}(\omega) = \mathbb{F}_R(\omega) + \mathrm{i}\mathbb{F}_I(\omega)}$$

$$= \mathbb{F}_R(ck)\sin(ckt_0) - \mathbb{F}_I(ck)\cos(ckt_0). \tag{16}$$

Considering that $F(t) * Q_k(t)|_{t=t_0} = (2\pi)^{-1}\int_{-\infty}^{\infty}\mathbb{F}(\omega)\mathbb{Q}_k(\omega)\exp(-\mathrm{i}\omega t_0)\,\mathrm{d}\omega$, we find

$$\int_{-\infty}^{\infty}\mathbb{F}(\omega)\mathbb{Q}_k(\omega)\exp(-\mathrm{i}\omega t_0)\,\mathrm{d}\omega = 2\pi[\mathbb{F}_R(ck)\sin(ckt_0) - \mathbb{F}_I(ck)\cos(ckt_0)]. \tag{17}$$

By the same token, if $F(t)$ happens to be purely imaginary — in which case $\mathbb{F}(\omega)$ will be anti-Hermitian — we will find

$$\int_{-\infty}^{\infty}\mathbb{F}(\omega)\mathbb{Q}_k(\omega)\exp(-\mathrm{i}\omega t_0)\,\mathrm{d}\omega = 2\pi\mathrm{i}[\mathbb{F}_R(ck)\cos(ckt_0) + \mathbb{F}_I(ck)\sin(ckt_0)]. \tag{18}$$

The electric and magnetic fields of Eqs.(13a) and (13b), evaluated at $t_0 > T$, may now be written as follows:

$$\boldsymbol{E}(\boldsymbol{r},t_0) = \frac{c}{2\pi^2\varepsilon_0}\int_{-\infty}^{\infty}\left[\frac{\sin(kR) - kR\cos(kR)}{k^4}\right]\exp(\mathrm{i}\boldsymbol{k}\cdot\boldsymbol{r})$$
$$\times\left\{\left(k^2\vec{\mathbb{P}}_R(ck) - [\vec{\mathbb{P}}_R(ck)\cdot\boldsymbol{k}]\boldsymbol{k}\right)\sin(ckt_0) - \left(k^2\vec{\mathbb{P}}_I(ck) - [\vec{\mathbb{P}}_I(ck)\cdot\boldsymbol{k}]\boldsymbol{k}\right)\cos(ckt_0)\right\}\mathrm{d}\boldsymbol{k}. \tag{19a}$$

$$\boldsymbol{H}(\boldsymbol{r},t_0) = \frac{\mathrm{i}c^2}{2\pi^2}\int_{-\infty}^{\infty}\left[\frac{\sin(kR) - kR\cos(kR)}{k^3}\right]\exp(\mathrm{i}\boldsymbol{k}\cdot\boldsymbol{r})\,\boldsymbol{k}\times[\vec{\mathbb{P}}_R(ck)\cos(ckt_0) + \vec{\mathbb{P}}_I(ck)\sin(ckt_0)]\mathrm{d}\boldsymbol{k}. \tag{19b}$$

These fields will now be used to compute the EM energy, linear momentum, and angular momentum of the radiated fields.

**4. Radiated energy**. The total radiated energy is the integral over all space of the $E$-field energy-density $\tfrac{1}{2}\varepsilon_0\boldsymbol{E}(\boldsymbol{r},t_0)\cdot\boldsymbol{E}(\boldsymbol{r},t_0)$, plus the $H$-field energy-density $\tfrac{1}{2}\mu_0\boldsymbol{H}(\boldsymbol{r},t_0)\cdot\boldsymbol{H}(\boldsymbol{r},t_0)$. As shown in Appendix C, the total radiated energy, which is independent of $t_0$ (assuming $t_0 > T$, the excitation duration), is given by

$$\mathcal{E}(t_0) = \frac{8R^6c^2}{3\varepsilon_0}\int_{k=0}^{\infty}\left[\frac{\sin(kR) - kR\cos(kR)}{(kR)^3}\right]^2 k^4\vec{\mathbb{P}}(ck)\cdot\vec{\mathbb{P}}^*(ck)\mathrm{d}k. \tag{20}$$

Considering that the radius $R$ of the spherical dipole is much less than the radiated wavelengths, that is, $kR \ll 1$, Eq.(20), upon further simplification, yields

$$\mathcal{E}(t_0) \cong \frac{8R^6c^2}{27\varepsilon_0}\int_0^{\infty}k^4\vec{\mathbb{P}}(ck)\cdot\vec{\mathbb{P}}^*(ck)\mathrm{d}k \cong \frac{8\mu_0R^6}{27c}\int_0^{\infty}\omega^4\vec{\mathbb{P}}(\omega)\cdot\vec{\mathbb{P}}^*(\omega)\mathrm{d}\omega. \tag{21}$$

The total EM field energy at $t_0$ is, of course, the sum of the incident energy given (per unit cross-sectional area) in Eq.(2), the radiated energy given in Eqs.(20) or (21), and the cross-term arising from the interference between the incident and radiated wave packets. For the incident packet given in Eq.(1), the cross-term, as shown in Appendix C, will be

$$\mathcal{E}_{\text{cross}}(t_0) = -2\varepsilon_0R^3\int_{-\infty}^{\infty}\left[\frac{\sin(k_zR) - k_zR\cos(k_zR)}{(k_zR)^3}\right]k_z\chi''(ck_z)\vec{\mathbb{E}}^{(\text{inc})}(k_z)\cdot\vec{\mathbb{E}}^{(\text{inc})*}(k_z)\mathrm{d}k_z. \tag{22}$$

In accordance with Eq.(22), the imaginary part $\chi''$ of the susceptibility accounts for the energy that is taken away from the incident wave-packet; see Eq.(9). A fraction of this energy is (permanently) absorbed within the particle when the damping coefficient $\gamma \neq 0$; the rest of the energy is (temporarily) picked up by the dipole and, subsequently, scattered away.

**5. Radiated linear momentum**. The EM field radiated by the spherical dipole does not carry any linear momentum. This is shown in Appendix D, where it is also confirmed that the EM force exerted by the dipole radiation on the spherical particle is zero. There is, however, a non-zero cross-term in the expression of the total EM linear momentum which is independent of time provided that $t_0 > T$. For the one-dimensional light pulse described in Eqs.(1), the cross-term, computed in Appendix D, is found to be

$$\boldsymbol{\wp}_{\text{cross}}(t_0) = -\frac{2R^3\hat{\boldsymbol{z}}}{c^2 Z_0}\int_{-\infty}^{\infty}\left[\frac{\sin(R\omega/c)-(R\omega/c)\cos(R\omega/c)}{(R\omega/c)^3}\right]\text{Im}[\omega\chi(\omega)]\vec{\mathbb{E}}^{(\text{inc})}(\omega)\cdot\vec{\mathbb{E}}^{(\text{inc})*}(\omega)\mathrm{d}\omega. \quad (23)$$

When the radius $R$ of the particle is sufficiently small, the above equation is simplified, yielding

$$\boldsymbol{\wp}_{\text{cross}}(t_0) \cong -\frac{(4\pi/3)R^3\hat{\boldsymbol{z}}}{\pi c^2 Z_0}\int_0^{\infty}\text{Im}[\omega\chi(\omega)]\vec{\mathbb{E}}^{(\text{inc})}(\omega)\cdot\vec{\mathbb{E}}^{(\text{inc})*}(\omega)\mathrm{d}\omega. \quad (24)$$

The momentum in Eq.(23) is deducted from the EM momentum of the incident packet. Conservation of momentum then dictates that this same momentum must have been picked up (as mechanical momentum) by the particle. Some of this momentum transfer is due to absorption (assuming the damping coefficient $\gamma$ of the material is non-zero); the rest is transferred to the particle in consequence of scattering.

**6. Radiated angular momentum**. In contrast to linear momentum, the oscillating spherical dipole is capable of radiating EM angular momentum. Detailed calculations pertaining to the radiated angular momentum are given in Appendix E. Assuming the one-dimensional incident light pulse described by Eqs.(1), and assuming $\vec{\mathbb{P}}(\omega) = \mathbb{P}_R(\omega)\hat{\boldsymbol{x}} + \mathrm{i}\mathbb{P}_I(\omega)\hat{\boldsymbol{y}}$.The total radiated angular momentum is found to be

$$\boldsymbol{\mathcal{L}}(t_0) = \frac{16\mu_0 R^6\hat{\boldsymbol{z}}}{3c}\int_0^{\infty}\left[\frac{\sin(R\omega/c)-(R\omega/c)\cos(R\omega/c)}{(R\omega/c)^3}\right]^2\omega^3\mathbb{P}_R(\omega)\mathbb{P}_I(\omega)\mathrm{d}\omega. \quad (25)$$

For a particle having a sufficiently small radius $R$, the above equation simplifies, yielding

$$\boldsymbol{\mathcal{L}}(t_0) \cong \frac{16\mu_0 R^6\hat{\boldsymbol{z}}}{27c}\int_0^{\infty}\omega^3\mathbb{P}_R(\omega)\mathbb{P}_I(\omega)\mathrm{d}\omega. \quad (26)$$

Also computed in Appendix E is the cross-term between the radiated field and the incident light pulse, which turns out to be

$$\boldsymbol{\mathcal{L}}_{\text{cross}}(t_0) = -2(\varepsilon_0 R^3/c)\int_{-\infty}^{\infty}\left[\frac{\sin(k_z R)-k_z R\cos(k_z R)}{(k_z R)^3}\right]\chi(ck_z)\vec{\mathbb{E}}^{(\text{inc})}(k_z)\times\vec{\mathbb{E}}^{(\text{inc})*}(k_z)\mathrm{d}k_z. \quad (27)$$

If the particle happens to absorb some of the incident light, then a fraction of the incident angular momentum will be transferred to the particle in the form of mechanical angular momentum. In contrast, a transparent particle does *not* experience a net torque and, therefore, does not capture any angular momentum from the incident beam. In the latter case, the radiated angular momentum is what is removed from the incident wave-packet by way of interference between the radiated field and the incident field.

**7. Results and Discussion**. We now summarize the results of our calculations pertaining to the EM field radiated by the small spherical particle depicted in Fig.1 ("dipole radiation"), and also the computed energy as well as linear and angular momenta of the EM field that are contributed by the "cross-terms" arising from interference between the dipole radiation and the (unhindered) wave-packet, as the packet continues its propagation into the empty space beyond the particle.

**Case 1**: **Transparent dielectric particle**. In the case of a passive, non-absorptive dielectric particle, the EM energy of the dipole radiation is equal in magnitude and opposite in sign to the energy associated with the cross-terms in the region where the dipole radiation and the wave-packet overlap. The radiated field does not carry any linear momentum, but the cross-terms resulting from interference contribute to the overall momentum of the EM field. The EM momentum contributed by the cross-terms is equal in magnitude and opposite in sign to the mechanical momentum picked up by the particle during its encounter with the wave-packet. The transparent particle does not experience a net torque, indicating that the angular momentum content of the radiated dipolar field is cancelled out by the angular momentum contributed by the cross-terms of the interference pattern.

**Case 2: Absorptive particle**. Once again, the particle picks up some linear momentum from the wave-packet, but this time it could also take away some of the energy and angular momentum of the incident beam, converting the latter to mechanical angular momentum. As before, the directly radiated EM field carries some energy and angular momentum, but no linear momentum. The cross-terms arising in consequence of interference between the wave-packet and the dipole radiation contribute to the overall energy, linear momentum, and angular momentum of the EM field. However, these cross-terms do not fully cancel out the contributions of the dipolar field, because some energy now resides within the particle and, having experienced a net torque, the particle acquires some angular momentum as well. Needless to say, the particle also picks up some linear momentum, but this is similar, qualitatively speaking, to the case of a non-absorptive particle.

**Case 3**: **Pre-excited (or gainy) particle**. The susceptibility of the particle in the case of a gain medium has a negative oscillator strength $f$, which ensures that the EM energy of the dipole radiation plus the contribution to the energy by the cross-terms within the interference zone end up making a net positive contribution [see Eq.(9)]; this net addition to the EM energy is caused by the (stimulated) extraction of the stored energy of the pre-excited particle. The linear and angular momenta of the EM field will likewise receive contributions from the stored energy.

Our analysis of the present case sheds light on the stimulated emission process, considering that the stored energy of the particle primarily contributes to the cross-terms arising from interference between the dipole radiation and the (unhindered) wave-packet. It is also possible in this case to examine the effects of vacuum fluctuations on the emission process, thus drawing conclusions about the process of spontaneous emission. However, these and related issues are best left to the field of quantum electrodynamics [12].

## Appendix A

### Incident beam of uniform cross-section in the $xy$-plane propagating along the $z$-axis

The incident light pulse of finite duration $T$, having a finite width $cT$ along the $z$-axis, infinite extent in the $xy$-plane, and an arbitrary polarization state with its $E$-field $\boldsymbol{E}^{(\mathrm{i})}(\boldsymbol{r},t)=E_x^{(\mathrm{i})}(z,t)\hat{\boldsymbol{x}}+E_y^{(\mathrm{i})}(z,t)\hat{\boldsymbol{y}}$ confined to the $xy$-plane, travels along the $z$-axis. The $E$ and $H$ fields may thus be expressed as superpositions of plane-waves of various frequencies $\omega$ and propagation vectors $\boldsymbol{k}=k_z\hat{\boldsymbol{z}}=(\omega/c)\hat{\boldsymbol{z}}$, as follows:

$$\boldsymbol{E}^{(\mathrm{i})}(\boldsymbol{r},t)=(2\pi)^{-1}\int_{-\infty}^{\infty}\vec{\mathbb{E}}^{(\mathrm{i})}(\omega)\exp[\mathrm{i}(\omega/c)z-\mathrm{i}\omega t]\mathrm{d}\omega, \tag{A1}$$

$$\boldsymbol{H}^{(\mathrm{i})}(\boldsymbol{r},t)=(2\pi Z_0)^{-1}\int_{-\infty}^{\infty}\hat{\boldsymbol{z}}\times\vec{\mathbb{E}}^{(\mathrm{i})}(\omega)\exp[\mathrm{i}(\omega/c)z-\mathrm{i}\omega t]\mathrm{d}\omega. \tag{A2}$$

Here $c=1/\sqrt{\mu_0\varepsilon_0}$ is the speed of light in vacuum, while $Z_0=\sqrt{\mu_0/\varepsilon_0}$ is the impedance of free space. The energy content per unit cross-sectional area of the beam, $\mathcal{E}^{(\mathrm{i})}(t)$, is readily computed by integrating the $E$- and $H$-field energy densities over the entire $z$-axis. In the absence of losses—which is characteristic of propagation in vacuum—the beam's energy content is found to be time-independent.

$$\begin{aligned}\mathcal{E}^{(\mathrm{i})}(t)&=½\varepsilon_0\int_{-\infty}^{\infty}\boldsymbol{E}^{(\mathrm{i})}(\boldsymbol{r},t)\cdot\boldsymbol{E}^{(\mathrm{i})}(\boldsymbol{r},t)\mathrm{d}z+½\mu_0\int_{-\infty}^{\infty}\boldsymbol{H}^{(\mathrm{i})}(\boldsymbol{r},t)\cdot\boldsymbol{H}^{(\mathrm{i})}(\boldsymbol{r},t)\mathrm{d}z\\
&=\frac{\varepsilon_0}{4\pi^2}\iint_{-\infty}^{\infty}\vec{\mathbb{E}}^{(\mathrm{i})}(\omega)\cdot\vec{\mathbb{E}}^{(\mathrm{i})}(\omega')\exp[-\mathrm{i}(\omega+\omega')t]\int_{-\infty}^{\infty}\exp[\mathrm{i}(\omega+\omega')z/c]\,\mathrm{d}z\mathrm{d}\omega\mathrm{d}\omega'\\
&=\frac{\varepsilon_0}{2\pi}\iint_{-\infty}^{\infty}\vec{\mathbb{E}}^{(\mathrm{i})}(\omega)\cdot\vec{\mathbb{E}}^{(\mathrm{i})}(\omega')\exp[-\mathrm{i}(\omega+\omega')t]\,\delta[(\omega+\omega')/c]\,\mathrm{d}\omega\mathrm{d}\omega'\\
&=\frac{1}{2\pi Z_0}\int_{-\infty}^{\infty}\vec{\mathbb{E}}^{(\mathrm{i})}(\omega)\cdot\vec{\mathbb{E}}^{(\mathrm{i})*}(\omega)\mathrm{d}\omega. \qquad \boxed{\int_{-\infty}^{\infty}\exp(\mathrm{i}\omega z)\,\mathrm{d}z=2\pi\delta(\omega)}\end{aligned} \tag{A3}$$

The linear momentum per unit cross-sectional area of the beam is similarly evaluated, as follows:

$$\begin{aligned}\boldsymbol{\wp}^{(\mathrm{i})}(t)&=c^{-2}\int_{-\infty}^{\infty}\boldsymbol{E}^{(\mathrm{i})}(\boldsymbol{r},t)\times\boldsymbol{H}^{(\mathrm{i})}(\boldsymbol{r},t)\mathrm{d}z\\
&=\frac{\hat{\boldsymbol{z}}}{4\pi^2c^2Z_0}\iint_{-\infty}^{\infty}\vec{\mathbb{E}}^{(\mathrm{i})}(\omega)\cdot\vec{\mathbb{E}}^{(\mathrm{i})}(\omega')\exp[-\mathrm{i}(\omega+\omega')t]\int_{-\infty}^{\infty}\exp[\mathrm{i}(\omega+\omega')z/c]\,\mathrm{d}z\mathrm{d}\omega\mathrm{d}\omega'\\
&=\frac{\hat{\boldsymbol{z}}}{2\pi cZ_0}\iint_{-\infty}^{\infty}\vec{\mathbb{E}}^{(\mathrm{i})}(\omega)\cdot\vec{\mathbb{E}}^{(\mathrm{i})}(\omega')\exp[-\mathrm{i}(\omega+\omega')t]\,\delta(\omega+\omega')\,\mathrm{d}\omega\mathrm{d}\omega'\\
&=(\varepsilon_0/2\pi)\hat{\boldsymbol{z}}\int_{-\infty}^{\infty}\vec{\mathbb{E}}^{(\mathrm{i})}(\omega)\cdot\vec{\mathbb{E}}^{(\mathrm{i})*}(\omega)\mathrm{d}\omega=\left[\mathcal{E}^{(\mathrm{i})}(t)/c\right]\hat{\boldsymbol{z}}.\end{aligned} \tag{A4}$$

As expected, the overall energy and momentum content of the beam are seen to be time-independent. Computing the beam's angular momentum requires that we pay close attention to the structural features of the light pulse, because the uniform cross-section of the beam described by Eqs.(A1) and (A2) hides its angular momentum content, as revealed by the following calculation:

$$\begin{aligned}\boldsymbol{\mathcal{L}}^{(\mathrm{i})}(t)&=c^{-2}\int_{-\infty}^{\infty}\boldsymbol{r}\times\left[\boldsymbol{E}^{(\mathrm{i})}(\boldsymbol{r},t)\times\boldsymbol{H}^{(\mathrm{i})}(\boldsymbol{r},t)\right]\mathrm{d}z\\
&=\frac{1}{4\pi^2c^2Z_0}\iint_{-\infty}^{\infty}\vec{\mathbb{E}}^{(\mathrm{i})}(\omega)\cdot\vec{\mathbb{E}}^{(\mathrm{i})}(\omega')\exp[-\mathrm{i}(\omega+\omega')t]\int_{-\infty}^{\infty}\boldsymbol{r}\times\hat{\boldsymbol{z}}\exp[\mathrm{i}(\omega+\omega')z/c]\,\mathrm{d}z\mathrm{d}\omega\mathrm{d}\omega'\\
&=\frac{y\hat{\boldsymbol{x}}-x\hat{\boldsymbol{y}}}{2\pi cZ_0}\iint_{-\infty}^{\infty}\vec{\mathbb{E}}^{(\mathrm{i})}(\omega)\cdot\vec{\mathbb{E}}^{(\mathrm{i})}(\omega')\exp[-\mathrm{i}(\omega+\omega')t]\,\delta(\omega+\omega')\mathrm{d}\omega\mathrm{d}\omega'\\
&=(\varepsilon_0/2\pi)(x\hat{\boldsymbol{x}}+y\hat{\boldsymbol{y}})\times\hat{\boldsymbol{z}}\int_{-\infty}^{\infty}\vec{\mathbb{E}}^{(\mathrm{i})}(\omega)\cdot\vec{\mathbb{E}}^{(\mathrm{i})*}(\omega)\mathrm{d}\omega.\end{aligned} \tag{A5}$$

Considering that the integral of $(x\hat{\boldsymbol{x}}+y\hat{\boldsymbol{y}})\times\hat{\boldsymbol{z}}$ around any circle in the $xy$-plane vanishes, our one-dimensional wave-packet is *inadequate* for our angular momentum calculations. As shown in Appendix B, computing the EM angular momentum requires a wave-packet whose cross-sectional area is finite.

## Appendix B

### Finite duration, finite cross-sectional area wave-packet propagating in free space along the *z*-axis

Expanding the electric and magnetic fields of a compact wave-packet in $(k_x, k_y, k_z)$ space, we will have

$$\boldsymbol{E}(\boldsymbol{r},t) = \frac{1}{(2\pi)^3}\int \vec{\mathbb{E}}(\boldsymbol{k}) \exp[\mathrm{i}(\boldsymbol{k}\cdot\boldsymbol{r} - \omega t)]\,\mathrm{d}\boldsymbol{k}. \tag{B1a}$$

$$\boldsymbol{H}(\boldsymbol{r},t) = \frac{1}{(2\pi)^3 Z_0}\int \hat{\boldsymbol{k}} \times \vec{\mathbb{E}}(\boldsymbol{k}) \exp[\mathrm{i}(\boldsymbol{k}\cdot\boldsymbol{r} - \omega t)]\,\mathrm{d}\boldsymbol{k}. \tag{B1b}$$

In the above equations

$$\boldsymbol{k} = k_x\hat{\boldsymbol{x}} + k_y\hat{\boldsymbol{y}} + k_z\hat{\boldsymbol{z}}, \tag{B2a}$$

$$\omega = ck_z\sqrt{1 + (k_x/k_z)^2 + (k_y/k_z)^2}, \tag{B2b}$$

$$\hat{\boldsymbol{k}} = c\boldsymbol{k}/\omega = (k_x\hat{\boldsymbol{x}} + k_y\hat{\boldsymbol{y}} + k_z\hat{\boldsymbol{z}})/\Big[k_z\sqrt{1 + (k_x/k_z)^2 + (k_y/k_z)^2}\,\Big], \tag{B2c}$$

$$\partial\omega/\partial k_x = c^2 k_x/\omega, \tag{B2d}$$

$$\partial\omega/\partial k_y = c^2 k_y/\omega, \tag{B2e}$$

$$\partial\omega/\partial k_z = c^2 k_z/\omega. \tag{B2f}$$

$$\partial_{k_x}\hat{\boldsymbol{k}} = \partial_{k_x}(c\boldsymbol{k}/\omega) = -(c/\omega^2)(\partial_{k_x}\omega)\boldsymbol{k} + (c/\omega)(\partial_{k_x}\boldsymbol{k}) = -(c/\omega)^3 k_x\boldsymbol{k} + (c/\omega)\hat{\boldsymbol{x}}. \tag{B2g}$$

The total energy content of the packet is found to be

$$\begin{aligned}
\mathcal{E}(t) &= \tfrac{1}{2}\varepsilon_0\int \boldsymbol{E}(\boldsymbol{r},t)\cdot\boldsymbol{E}(\boldsymbol{r},t)\mathrm{d}\boldsymbol{r} + \tfrac{1}{2}\mu_0\int \boldsymbol{H}(\boldsymbol{r},t)\cdot\boldsymbol{H}(\boldsymbol{r},t)\mathrm{d}\boldsymbol{r}\\
&= \frac{1}{2(2\pi)^6}\int\big[\varepsilon_0\vec{\mathbb{E}}(\boldsymbol{k})\cdot\vec{\mathbb{E}}(\boldsymbol{k}') + \mu_0\vec{\mathbb{H}}(\boldsymbol{k})\cdot\vec{\mathbb{H}}(\boldsymbol{k}')\big]\exp[\mathrm{i}(\boldsymbol{k}+\boldsymbol{k}')\cdot\boldsymbol{r}]\exp[-\mathrm{i}(\omega+\omega')t]\,\mathrm{d}\boldsymbol{k}\mathrm{d}\boldsymbol{k}'\mathrm{d}\boldsymbol{r}\\
&= \frac{1}{2(2\pi)^3}\int\big[\varepsilon_0\vec{\mathbb{E}}(\boldsymbol{k})\cdot\vec{\mathbb{E}}(\boldsymbol{k}') + \mu_0\vec{\mathbb{H}}(\boldsymbol{k})\cdot\vec{\mathbb{H}}(\boldsymbol{k}')\big]\exp[-\mathrm{i}(\omega+\omega')t]\\
&\qquad\qquad\times\delta(k_x+k_x')\,\delta(k_y+k_y')\,\delta(k_z+k_z')\,\mathrm{d}\boldsymbol{k}\mathrm{d}\boldsymbol{k}'\\
&= \frac{1}{2(2\pi)^3}\int\big[\varepsilon_0\vec{\mathbb{E}}(\boldsymbol{k})\cdot\vec{\mathbb{E}}^*(\boldsymbol{k}) + \mu_0\vec{\mathbb{H}}(\boldsymbol{k})\cdot\vec{\mathbb{H}}^*(\boldsymbol{k})\big]\mathrm{d}\boldsymbol{k}\\
&= \frac{1}{2(2\pi)^3}\int\big\{\varepsilon_0\vec{\mathbb{E}}(\boldsymbol{k})\cdot\vec{\mathbb{E}}^*(\boldsymbol{k}) + (\mu_0/Z_0^2)[\hat{\boldsymbol{k}}\times\vec{\mathbb{E}}(\boldsymbol{k})]\cdot[\hat{\boldsymbol{k}}\times\vec{\mathbb{E}}^*(\boldsymbol{k})]\big\}\mathrm{d}\boldsymbol{k}\\
&= \frac{\varepsilon_0}{2(2\pi)^3}\int\big\{\vec{\mathbb{E}}(\boldsymbol{k})\cdot\vec{\mathbb{E}}^*(\boldsymbol{k}) + (\hat{\boldsymbol{k}}\cdot\hat{\boldsymbol{k}})\big[\vec{\mathbb{E}}(\boldsymbol{k})\cdot\vec{\mathbb{E}}^*(\boldsymbol{k})\big] - \cancelto{0}{[\hat{\boldsymbol{k}}\cdot\vec{\mathbb{E}}^*(\boldsymbol{k})]}\cancelto{0}{[\hat{\boldsymbol{k}}\cdot\vec{\mathbb{E}}(\boldsymbol{k})]}\big\}\mathrm{d}\boldsymbol{k}\\
&= \frac{\varepsilon_0}{(2\pi)^3}\int\vec{\mathbb{E}}(\boldsymbol{k})\cdot\vec{\mathbb{E}}^*(\boldsymbol{k})\mathrm{d}\boldsymbol{k}.
\end{aligned}\tag{B3}$$

Similarly, the linear momentum of the packet is readily computed as follows:

$$\begin{aligned}
\boldsymbol{\mathscr{p}}(t) &= (1/c^2)\int\boldsymbol{E}(\boldsymbol{r},t)\times\boldsymbol{H}(\boldsymbol{r},t)\mathrm{d}\boldsymbol{r}\\
&= \frac{1}{(2\pi)^6 Z_0 c^2}\int\vec{\mathbb{E}}(\boldsymbol{k}')\times[\hat{\boldsymbol{k}}\times\vec{\mathbb{E}}(\boldsymbol{k})]\exp[\mathrm{i}(\boldsymbol{k}+\boldsymbol{k}')\cdot\boldsymbol{r}]\exp[-\mathrm{i}(\omega+\omega')t]\,\mathrm{d}\boldsymbol{k}\mathrm{d}\boldsymbol{k}'\mathrm{d}\boldsymbol{r}\\
&= \frac{1}{(2\pi)^3 Z_0 c^2}\int\big\{[\vec{\mathbb{E}}(\boldsymbol{k}')\cdot\vec{\mathbb{E}}(\boldsymbol{k})]\hat{\boldsymbol{k}} - [\vec{\mathbb{E}}(\boldsymbol{k}')\cdot\hat{\boldsymbol{k}}]\vec{\mathbb{E}}(\boldsymbol{k})\big\}\exp[-\mathrm{i}(\omega+\omega')t]\\
&\qquad\qquad\times\delta(k_x+k_x')\,\delta(k_y+k_y')\,\delta(k_z+k_z')\,\mathrm{d}\boldsymbol{k}\mathrm{d}\boldsymbol{k}'\\
&= \frac{\varepsilon_0}{(2\pi)^3 c}\int\big\{[\vec{\mathbb{E}}^*(\boldsymbol{k})\cdot\vec{\mathbb{E}}(\boldsymbol{k})]\hat{\boldsymbol{k}} - \cancelto{0}{[\vec{\mathbb{E}}^*(\boldsymbol{k})\cdot\hat{\boldsymbol{k}}]}\vec{\mathbb{E}}(\boldsymbol{k})\big\}\mathrm{d}\boldsymbol{k} = \frac{\varepsilon_0}{(2\pi)^3 c}\int[\vec{\mathbb{E}}(\boldsymbol{k})\cdot\vec{\mathbb{E}}^*(\boldsymbol{k})]\hat{\boldsymbol{k}}\mathrm{d}\boldsymbol{k}.
\end{aligned}\tag{B4}$$

Computation of the angular momentum is somewhat more complicated, involving the following steps:

$$\boldsymbol{\mathcal{L}}(t)=(1/c^2)\int\boldsymbol{r}\times[\boldsymbol{E}(\boldsymbol{r},t)\times\boldsymbol{H}(\boldsymbol{r},t)]\mathrm{d}\boldsymbol{r}$$

$$=\frac{1}{(2\pi)^6 Z_0 c^2}\int\boldsymbol{r}\times\left\{\vec{\mathbb{E}}(\boldsymbol{k})\times[\hat{\boldsymbol{k}}'\times\vec{\mathbb{E}}(\boldsymbol{k}')]\right\}\exp[\mathrm{i}(\boldsymbol{k}+\boldsymbol{k}')\cdot\boldsymbol{r}]\exp(-\mathrm{i}\omega t)\exp(-\mathrm{i}\omega' t)\,\mathrm{d}\boldsymbol{k}\mathrm{d}\boldsymbol{k}'\mathrm{d}\boldsymbol{r}$$

$$=\frac{\varepsilon_0}{(2\pi)^6 c}\int\{[\vec{\mathbb{E}}(\boldsymbol{k})\cdot\vec{\mathbb{E}}(\boldsymbol{k}')](\boldsymbol{r}\times\hat{\boldsymbol{k}}')-[\vec{\mathbb{E}}(\boldsymbol{k})\cdot\hat{\boldsymbol{k}}'][\boldsymbol{r}\times\vec{\mathbb{E}}(\boldsymbol{k}')]\}\exp[\mathrm{i}(\boldsymbol{k}+\boldsymbol{k}')\cdot\boldsymbol{r}]\exp(-\mathrm{i}\omega t)\exp(-\mathrm{i}\omega' t)\,\mathrm{d}\boldsymbol{k}\mathrm{d}\boldsymbol{k}'\mathrm{d}\boldsymbol{r}$$

$$=\frac{\varepsilon_0}{(2\pi)^6 c}\int\{[\vec{\mathbb{E}}(\boldsymbol{k})\cdot\vec{\mathbb{E}}(\boldsymbol{k}')][(y\hat{k}_z'-z\hat{k}_y')\hat{\boldsymbol{x}}+(z\hat{k}_x'-x\hat{k}_z')\hat{\boldsymbol{y}}+(x\hat{k}_y'-y\hat{k}_x')\hat{\boldsymbol{z}}]$$

$$-[\vec{\mathbb{E}}(\boldsymbol{k})\cdot\hat{\boldsymbol{k}}']\{[y\mathbb{E}_z(\boldsymbol{k}')-z\mathbb{E}_y(\boldsymbol{k}')]\hat{\boldsymbol{x}}+[z\mathbb{E}_x(\boldsymbol{k}')-x\mathbb{E}_z(\boldsymbol{k}')]\hat{\boldsymbol{y}}+[x\mathbb{E}_y(\boldsymbol{k}')-y\mathbb{E}_x(\boldsymbol{k}')]\hat{\boldsymbol{z}}\}\Big\}$$

$$\times\exp[\mathrm{i}(\boldsymbol{k}+\boldsymbol{k}')\cdot\boldsymbol{r}]\exp(-\mathrm{i}\omega t)\exp(-\mathrm{i}\omega' t)\,\mathrm{d}\boldsymbol{k}\mathrm{d}\boldsymbol{k}'\mathrm{d}\boldsymbol{r}$$

$$=\frac{\varepsilon_0}{(2\pi)^6 c}\int\Big\{x\{[\vec{\mathbb{E}}(\boldsymbol{k})\cdot\vec{\mathbb{E}}(\boldsymbol{k}')](\hat{k}_y'\hat{\boldsymbol{z}}-\hat{k}_z'\hat{\boldsymbol{y}})-[\vec{\mathbb{E}}(\boldsymbol{k})\cdot\hat{\boldsymbol{k}}'][\mathbb{E}_y(\boldsymbol{k}')\hat{\boldsymbol{z}}-\mathbb{E}_z(\boldsymbol{k}')\hat{\boldsymbol{y}}]\}$$

$$+y\{[\vec{\mathbb{E}}(\boldsymbol{k})\cdot\vec{\mathbb{E}}(\boldsymbol{k}')](\hat{k}_z'\hat{\boldsymbol{x}}-\hat{k}_x'\hat{\boldsymbol{z}})-[\vec{\mathbb{E}}(\boldsymbol{k})\cdot\hat{\boldsymbol{k}}'][\mathbb{E}_z(\boldsymbol{k}')\hat{\boldsymbol{x}}-\mathbb{E}_x(\boldsymbol{k}')\hat{\boldsymbol{z}}]\}$$

$$+z\{[\vec{\mathbb{E}}(\boldsymbol{k})\cdot\vec{\mathbb{E}}(\boldsymbol{k}')](\hat{k}_x'\hat{\boldsymbol{y}}-\hat{k}_y'\hat{\boldsymbol{x}})-[\vec{\mathbb{E}}(\boldsymbol{k})\cdot\hat{\boldsymbol{k}}'][\mathbb{E}_x(\boldsymbol{k}')\hat{\boldsymbol{y}}-\mathbb{E}_y(\boldsymbol{k}')\hat{\boldsymbol{x}}]\}\Big\}$$

$$\times\exp[\mathrm{i}(\boldsymbol{k}+\boldsymbol{k}')\cdot\boldsymbol{r}]\exp(-\mathrm{i}\omega t)\exp(-\mathrm{i}\omega' t)\,\mathrm{d}\boldsymbol{k}\mathrm{d}\boldsymbol{k}'\mathrm{d}\boldsymbol{r}$$

$$=-\frac{\mathrm{i}\varepsilon_0}{(2\pi)^3 c}\int\Big\{\delta'(k_x+k_x')\delta(k_y+k_y')\delta(k_z+k_z')\{[\vec{\mathbb{E}}(\boldsymbol{k})\cdot\vec{\mathbb{E}}(\boldsymbol{k}')](\hat{k}_y'\hat{\boldsymbol{z}}-\hat{k}_z'\hat{\boldsymbol{y}})-[\vec{\mathbb{E}}(\boldsymbol{k})\cdot\hat{\boldsymbol{k}}'][\mathbb{E}_y(\boldsymbol{k}')\hat{\boldsymbol{z}}-\mathbb{E}_z(\boldsymbol{k}')\hat{\boldsymbol{y}}]\}$$

$$+\delta(k_x+k_x')\delta'(k_y+k_y')\delta(k_z+k_z')\{[\vec{\mathbb{E}}(\boldsymbol{k})\cdot\vec{\mathbb{E}}(\boldsymbol{k}')](\hat{k}_z'\hat{\boldsymbol{x}}-\hat{k}_x'\hat{\boldsymbol{z}})-[\vec{\mathbb{E}}(\boldsymbol{k})\cdot\hat{\boldsymbol{k}}'][\mathbb{E}_z(\boldsymbol{k}')\hat{\boldsymbol{x}}-\mathbb{E}_x(\boldsymbol{k}')\hat{\boldsymbol{z}}]\}$$

$$+\delta(k_x+k_x')\delta(k_y+k_y')\delta'(k_z+k_z')\{[\vec{\mathbb{E}}(\boldsymbol{k})\cdot\vec{\mathbb{E}}(\boldsymbol{k}')](\hat{k}_x'\hat{\boldsymbol{y}}-\hat{k}_y'\hat{\boldsymbol{x}})-[\vec{\mathbb{E}}(\boldsymbol{k})\cdot\hat{\boldsymbol{k}}'][\mathbb{E}_x(\boldsymbol{k}')\hat{\boldsymbol{y}}-\mathbb{E}_y(\boldsymbol{k}')\hat{\boldsymbol{x}}]\}\Big\}$$

$$\times\exp(-\mathrm{i}\omega t)\exp(-\mathrm{i}\omega' t)\,\mathrm{d}\boldsymbol{k}\mathrm{d}\boldsymbol{k}'$$

$$=\frac{\mathrm{i}\varepsilon_0}{(2\pi)^3 c}\int\Big\{\{[\partial_{k_x}\vec{\mathbb{E}}(-\boldsymbol{k}')\cdot\vec{\mathbb{E}}(\boldsymbol{k}')](\hat{k}_y'\hat{\boldsymbol{z}}-\hat{k}_z'\hat{\boldsymbol{y}})-[\partial_{k_x}\vec{\mathbb{E}}(-\boldsymbol{k}')\cdot\hat{\boldsymbol{k}}'][\mathbb{E}_y(\boldsymbol{k}')\hat{\boldsymbol{z}}-\mathbb{E}_z(\boldsymbol{k}')\hat{\boldsymbol{y}}]\}$$

$$-\mathrm{i}(c^2k_x'/\omega')t\{[\vec{\mathbb{E}}(-\boldsymbol{k}')\cdot\vec{\mathbb{E}}(\boldsymbol{k}')](\hat{k}_y'\hat{\boldsymbol{z}}-\hat{k}_z'\hat{\boldsymbol{y}})-[\vec{\mathbb{E}}(-\boldsymbol{k}')\cdot\hat{\boldsymbol{k}}'][\mathbb{E}_y(\boldsymbol{k}')\hat{\boldsymbol{z}}-\mathbb{E}_z(\boldsymbol{k}')\hat{\boldsymbol{y}}]\}$$

$$+\{[\partial_{k_y}\vec{\mathbb{E}}(-\boldsymbol{k}')\cdot\vec{\mathbb{E}}(\boldsymbol{k}')](\hat{k}_z'\hat{\boldsymbol{x}}-\hat{k}_x'\hat{\boldsymbol{z}})-[\partial_{k_y}\vec{\mathbb{E}}(-\boldsymbol{k}')\cdot\hat{\boldsymbol{k}}'][\mathbb{E}_z(\boldsymbol{k}')\hat{\boldsymbol{x}}-\mathbb{E}_x(\boldsymbol{k}')\hat{\boldsymbol{z}}]\}$$

$$-\mathrm{i}(c^2k_y'/\omega')t\{[\vec{\mathbb{E}}(-\boldsymbol{k}')\cdot\vec{\mathbb{E}}(\boldsymbol{k}')](\hat{k}_z'\hat{\boldsymbol{x}}-\hat{k}_x'\hat{\boldsymbol{z}})-[\vec{\mathbb{E}}(-\boldsymbol{k}')\cdot\hat{\boldsymbol{k}}'][\mathbb{E}_z(\boldsymbol{k}')\hat{\boldsymbol{x}}-\mathbb{E}_x(\boldsymbol{k}')\hat{\boldsymbol{z}}]\}$$

$$+\{[\partial_{k_z}\vec{\mathbb{E}}(-\boldsymbol{k}')\cdot\vec{\mathbb{E}}(\boldsymbol{k}')](\hat{k}_x'\hat{\boldsymbol{y}}-\hat{k}_y'\hat{\boldsymbol{x}})-[\partial_{k_z}\vec{\mathbb{E}}(-\boldsymbol{k}')\cdot\hat{\boldsymbol{k}}'][\mathbb{E}_x(\boldsymbol{k}')\hat{\boldsymbol{y}}-\mathbb{E}_y(\boldsymbol{k}')\hat{\boldsymbol{x}}]\}\Big\}$$

$$-\mathrm{i}(c^2k_z'/\omega')t\{[\vec{\mathbb{E}}(-\boldsymbol{k}')\cdot\vec{\mathbb{E}}(\boldsymbol{k}')](\hat{k}_x'\hat{\boldsymbol{y}}-\hat{k}_y'\hat{\boldsymbol{x}})-[\vec{\mathbb{E}}(-\boldsymbol{k}')\cdot\hat{\boldsymbol{k}}'][\mathbb{E}_x(\boldsymbol{k}')\hat{\boldsymbol{y}}-\mathbb{E}_y(\boldsymbol{k}')\hat{\boldsymbol{x}}]\}\mathrm{d}\boldsymbol{k}'$$

$$=\frac{\mathrm{i}\varepsilon_0}{(2\pi)^3 c}\int\{[\vec{\mathbb{E}}(\boldsymbol{k})\cdot\partial_{k_x}\vec{\mathbb{E}}^*(\boldsymbol{k})](\hat{k}_y\hat{\boldsymbol{z}}-\hat{k}_z\hat{\boldsymbol{y}})-[\hat{\boldsymbol{k}}\cdot\partial_{k_x}\vec{\mathbb{E}}^*(\boldsymbol{k})][\mathbb{E}_y(\boldsymbol{k})\hat{\boldsymbol{z}}-\mathbb{E}_z(\boldsymbol{k})\hat{\boldsymbol{y}}]$$

$$+[\vec{\mathbb{E}}(\boldsymbol{k})\cdot\partial_{k_y}\vec{\mathbb{E}}^*(\boldsymbol{k})](\hat{k}_z\hat{\boldsymbol{x}}-\hat{k}_x\hat{\boldsymbol{z}})-[\hat{\boldsymbol{k}}\cdot\partial_{k_y}\vec{\mathbb{E}}^*(\boldsymbol{k})][\mathbb{E}_z(\boldsymbol{k})\hat{\boldsymbol{x}}-\mathbb{E}_x(\boldsymbol{k})\hat{\boldsymbol{z}}]$$

$$+[\vec{\mathbb{E}}(\boldsymbol{k})\cdot\partial_{k_z}\vec{\mathbb{E}}^*(\boldsymbol{k})](\hat{k}_x\hat{\boldsymbol{y}}-\hat{k}_y\hat{\boldsymbol{x}})-[\hat{\boldsymbol{k}}\cdot\partial_{k_z}\vec{\mathbb{E}}^*(\boldsymbol{k})][\mathbb{E}_x(\boldsymbol{k})\hat{\boldsymbol{y}}-\mathbb{E}_y(\boldsymbol{k})\hat{\boldsymbol{x}}]\}\mathrm{d}\boldsymbol{k}$$

$$=\frac{\mathrm{i}\varepsilon_0}{(2\pi)^3 c}\int\{\hat{\boldsymbol{x}}\times\{[\vec{\mathbb{E}}(\boldsymbol{k})\cdot\partial_{k_x}\vec{\mathbb{E}}^*(\boldsymbol{k})]\hat{\boldsymbol{k}}-[\hat{\boldsymbol{k}}\cdot\partial_{k_x}\vec{\mathbb{E}}^*(\boldsymbol{k})]\vec{\mathbb{E}}(\boldsymbol{k})\}+\hat{\boldsymbol{y}}\times\{[\vec{\mathbb{E}}(\boldsymbol{k})\cdot\partial_{k_y}\vec{\mathbb{E}}^*(\boldsymbol{k})]\hat{\boldsymbol{k}}-[\hat{\boldsymbol{k}}\cdot\partial_{k_y}\vec{\mathbb{E}}^*(\boldsymbol{k})]\vec{\mathbb{E}}(\boldsymbol{k})\}$$

$$+\hat{\boldsymbol{z}}\times\{[\vec{\mathbb{E}}(\boldsymbol{k})\cdot\partial_{k_z}\vec{\mathbb{E}}^*(\boldsymbol{k})]\hat{\boldsymbol{k}}-[\hat{\boldsymbol{k}}\cdot\partial_{k_z}\vec{\mathbb{E}}^*(\boldsymbol{k})]\vec{\mathbb{E}}(\boldsymbol{k})\}\Big\}\,\mathrm{d}\boldsymbol{k}$$

$\boldsymbol{A}\times(\boldsymbol{B}\times\boldsymbol{C})=(\boldsymbol{A}\cdot\boldsymbol{C})\boldsymbol{B}-(\boldsymbol{A}\cdot\boldsymbol{B})\boldsymbol{C}$

$$=\frac{\mathrm{i}\varepsilon_0}{(2\pi)^3 c}\int\Big\{\hat{\boldsymbol{x}}\times\{[\partial_{k_x}\vec{\mathbb{E}}^*(\boldsymbol{k})]\times[\hat{\boldsymbol{k}}\times\vec{\mathbb{E}}(\boldsymbol{k})]\}+\hat{\boldsymbol{y}}\times\{[\partial_{k_y}\vec{\mathbb{E}}^*(\boldsymbol{k})]\times[\hat{\boldsymbol{k}}\times\vec{\mathbb{E}}(\boldsymbol{k})]\}+\hat{\boldsymbol{z}}\times\{[\partial_{k_z}\vec{\mathbb{E}}^*(\boldsymbol{k})]\times[\hat{\boldsymbol{k}}\times\vec{\mathbb{E}}(\boldsymbol{k})]\}\Big\}\,\mathrm{d}\boldsymbol{k}$$

$$=\frac{\mathrm{i}\varepsilon_0}{(2\pi)^3 c}\int\Big\{\{[\hat{\boldsymbol{k}}\times\vec{\mathbb{E}}(\boldsymbol{k})]\cdot\boldsymbol{\nabla}_k\}\vec{\mathbb{E}}^*(\boldsymbol{k})-[\boldsymbol{\nabla}_k\cdot\vec{\mathbb{E}}^*(\boldsymbol{k})]\hat{\boldsymbol{k}}\times\vec{\mathbb{E}}(\boldsymbol{k})\Big\}\,\mathrm{d}\boldsymbol{k}. \qquad \text{(B5)}$$

The above integrand may be simplified using the identities $\boldsymbol{\nabla}\times\hat{\boldsymbol{k}}=0$ and $\hat{\boldsymbol{k}}\cdot\vec{\mathbb{E}}=\hat{\boldsymbol{k}}\cdot\vec{\mathbb{E}}^*=0$ and $\vec{\mathbb{E}}^*\times(\hat{\boldsymbol{k}}\times\vec{\mathbb{E}})=(\vec{\mathbb{E}}\cdot\vec{\mathbb{E}}^*)\hat{\boldsymbol{k}}$, the latter of which are rooted in Maxwell's equations. We also need to invoke standard vector identities to establish the following relations:

$$\boldsymbol{\nabla}\times(\boldsymbol{A}\times\boldsymbol{B})=(\boldsymbol{\nabla}\cdot\boldsymbol{B})\boldsymbol{A}-(\boldsymbol{\nabla}\cdot\boldsymbol{A})\boldsymbol{B}+(\boldsymbol{B}\cdot\boldsymbol{\nabla})\boldsymbol{A}-(\boldsymbol{A}\cdot\boldsymbol{\nabla})\boldsymbol{B}$$

$$\begin{aligned}\boldsymbol{\nabla}\times[\vec{\mathbb{E}}^*\times(\hat{\boldsymbol{k}}\times\vec{\mathbb{E}})]&=[\boldsymbol{\nabla}\cdot(\hat{\boldsymbol{k}}\times\vec{\mathbb{E}})]\vec{\mathbb{E}}^*-(\boldsymbol{\nabla}\cdot\vec{\mathbb{E}}^*)\hat{\boldsymbol{k}}\times\vec{\mathbb{E}}+[(\hat{\boldsymbol{k}}\times\vec{\mathbb{E}})\cdot\boldsymbol{\nabla}]\vec{\mathbb{E}}^*-(\vec{\mathbb{E}}^*\cdot\boldsymbol{\nabla})(\hat{\boldsymbol{k}}\times\vec{\mathbb{E}})\\&=-[\hat{\boldsymbol{k}}\cdot(\boldsymbol{\nabla}\times\vec{\mathbb{E}})]\vec{\mathbb{E}}^*-(\boldsymbol{\nabla}\cdot\vec{\mathbb{E}}^*)\hat{\boldsymbol{k}}\times\vec{\mathbb{E}}+[(\hat{\boldsymbol{k}}\times\vec{\mathbb{E}})\cdot\boldsymbol{\nabla}]\vec{\mathbb{E}}^*-\underbrace{[(\vec{\mathbb{E}}^*\cdot\boldsymbol{\nabla})\hat{\boldsymbol{k}}]}_{-(c/\omega)[(\vec{\mathbb{E}}^*\cdot\hat{\boldsymbol{k}})\hat{\boldsymbol{k}}-\vec{\mathbb{E}}^*]}\times\vec{\mathbb{E}}+[(\vec{\mathbb{E}}^*\cdot\boldsymbol{\nabla})\vec{\mathbb{E}}]\times\hat{\boldsymbol{k}}.\end{aligned}$$

$$\boldsymbol{\nabla}\cdot(\boldsymbol{A}\times\boldsymbol{B})=\boldsymbol{B}\cdot(\boldsymbol{\nabla}\times\boldsymbol{A})-\boldsymbol{A}\cdot(\boldsymbol{\nabla}\times\boldsymbol{B})\qquad\text{(B6)}$$

Considering that the left-hand side of Eq.(B6) equals $\boldsymbol{\nabla}\times[(\vec{\mathbb{E}}\cdot\vec{\mathbb{E}}^*)\hat{\boldsymbol{k}}]=[\boldsymbol{\nabla}(\vec{\mathbb{E}}\cdot\vec{\mathbb{E}}^*)]\times\hat{\boldsymbol{k}}$, we arrive at

$$[(\hat{\boldsymbol{k}}\times\vec{\mathbb{E}})\cdot\boldsymbol{\nabla}_k]\vec{\mathbb{E}}^*-(\boldsymbol{\nabla}_k\cdot\vec{\mathbb{E}}^*)\hat{\boldsymbol{k}}\times\vec{\mathbb{E}}=\big[\boldsymbol{\nabla}_k(\vec{\mathbb{E}}\cdot\vec{\mathbb{E}}^*)-(\vec{\mathbb{E}}^*\cdot\boldsymbol{\nabla}_k)\vec{\mathbb{E}}\big]\times\hat{\boldsymbol{k}}+\big[\hat{\boldsymbol{k}}\cdot(\boldsymbol{\nabla}_k\times\vec{\mathbb{E}})\big]\vec{\mathbb{E}}^*+(c/\omega)\vec{\mathbb{E}}^*\times\vec{\mathbb{E}}$$

$$\boldsymbol{\nabla}(\boldsymbol{A}\cdot\boldsymbol{B})=(\boldsymbol{A}\cdot\boldsymbol{\nabla})\boldsymbol{B}+(\boldsymbol{B}\cdot\boldsymbol{\nabla})\boldsymbol{A}+\boldsymbol{A}\times(\boldsymbol{\nabla}\times\boldsymbol{B})+\boldsymbol{B}\times(\boldsymbol{\nabla}\times\boldsymbol{A})$$

$$\begin{aligned}&=\big[(\vec{\mathbb{E}}\cdot\boldsymbol{\nabla}_{\boldsymbol{k}})\vec{\mathbb{E}}^*+\vec{\mathbb{E}}\times(\boldsymbol{\nabla}_{\boldsymbol{k}}\times\vec{\mathbb{E}}^*)+\vec{\mathbb{E}}^*\times\cancel{(\boldsymbol{\nabla}_{\boldsymbol{k}}\times\vec{\mathbb{E}})}\big]\times\hat{\boldsymbol{k}}+\cancel{\big[\hat{\boldsymbol{k}}\cdot(\boldsymbol{\nabla}_{\boldsymbol{k}}\times\vec{\mathbb{E}})\big]\vec{\mathbb{E}}^*}+\underbrace{(c/\omega)\vec{\mathbb{E}}^*\times\vec{\mathbb{E}}}_{\text{spin}}\\&=\big[(\vec{\mathbb{E}}\cdot\boldsymbol{\nabla}_{\boldsymbol{k}})\vec{\mathbb{E}}^*+\vec{\mathbb{E}}\times(\boldsymbol{\nabla}_{\boldsymbol{k}}\times\vec{\mathbb{E}}^*)\big]\times\hat{\boldsymbol{k}}+(c/\omega)\vec{\mathbb{E}}^*\times\vec{\mathbb{E}}.\end{aligned}\qquad\text{(B7)}$$

Aside from the spin component of the angular momentum, the remaining terms in the above expression are orthogonal to $\hat{\boldsymbol{k}}$ and can be further simplified, as follows:

$$\begin{aligned}(\vec{\mathbb{E}}\cdot\boldsymbol{\nabla}_{\boldsymbol{k}})\vec{\mathbb{E}}^*+\vec{\mathbb{E}}\times(\boldsymbol{\nabla}_{\boldsymbol{k}}\times\vec{\mathbb{E}}^*)&=\mathbb{E}_x\partial_{k_x}\vec{\mathbb{E}}^*+\mathbb{E}_y\partial_{k_y}\vec{\mathbb{E}}^*+\mathbb{E}_z\partial_{k_z}\vec{\mathbb{E}}^*+(\mathbb{E}_x\hat{\boldsymbol{x}}+\mathbb{E}_y\hat{\boldsymbol{y}}+\mathbb{E}_z\hat{\boldsymbol{z}})\\&\quad\times[(\partial_{k_y}\mathbb{E}_z^*-\partial_{k_z}\mathbb{E}_y^*)\hat{\boldsymbol{x}}+(\partial_{k_z}\mathbb{E}_x^*-\partial_{k_x}\mathbb{E}_z^*)\hat{\boldsymbol{y}}+(\partial_{k_x}\mathbb{E}_y^*-\partial_{k_y}\mathbb{E}_x^*)\hat{\boldsymbol{z}}]\\&=\big[\mathbb{E}_x\partial_{k_x}\mathbb{E}_x^*+\cancel{\mathbb{E}_y\partial_{k_y}\mathbb{E}_x^*}+\cancel{\mathbb{E}_z\partial_{k_z}\mathbb{E}_x^*}+\mathbb{E}_y(\partial_{k_x}\mathbb{E}_y^*-\cancel{\partial_{k_y}\mathbb{E}_x^*})-\mathbb{E}_z(\cancel{\partial_{k_z}\mathbb{E}_x^*}-\partial_{k_x}\mathbb{E}_z^*)\big]\hat{\boldsymbol{x}}\\&\quad+\big[\cancel{\mathbb{E}_x\partial_{k_x}\mathbb{E}_y^*}+\mathbb{E}_y\partial_{k_y}\mathbb{E}_y^*+\cancel{\mathbb{E}_z\partial_{k_z}\mathbb{E}_y^*}-\mathbb{E}_x(\cancel{\partial_{k_x}\mathbb{E}_y^*}-\partial_{k_y}\mathbb{E}_x^*)+\mathbb{E}_z(\partial_{k_y}\mathbb{E}_z^*-\cancel{\partial_{k_z}\mathbb{E}_y^*})\big]\hat{\boldsymbol{y}}\\&\quad+\big[\cancel{\mathbb{E}_x\partial_{k_x}\mathbb{E}_z^*}+\cancel{\mathbb{E}_y\partial_{k_y}\mathbb{E}_z^*}+\mathbb{E}_z\partial_{k_z}\mathbb{E}_z^*+\mathbb{E}_x(\partial_{k_z}\mathbb{E}_x^*-\cancel{\partial_{k_x}\mathbb{E}_z^*})-\mathbb{E}_y(\cancel{\partial_{k_y}\mathbb{E}_z^*}-\partial_{k_z}\mathbb{E}_y^*)\big]\hat{\boldsymbol{z}}\\&=(\vec{\mathbb{E}}\cdot\partial_{k_x}\vec{\mathbb{E}}^*)\hat{\boldsymbol{x}}+(\vec{\mathbb{E}}\cdot\partial_{k_y}\vec{\mathbb{E}}^*)\hat{\boldsymbol{y}}+(\vec{\mathbb{E}}\cdot\partial_{k_z}\vec{\mathbb{E}}^*)\hat{\boldsymbol{z}}.\end{aligned}\qquad\text{(B8)}$$

The total angular momentum of the wave-packet can finally be written in the following compact form:

$$\boldsymbol{\mathcal{L}}(t)=\frac{\mathrm{i}\varepsilon_0}{(2\pi)^3}\iiint_{-\infty}^{\infty}\omega^{-1}\{\big[(\vec{\mathbb{E}}\cdot\partial_{k_x}\vec{\mathbb{E}}^*)\hat{\boldsymbol{x}}+(\vec{\mathbb{E}}\cdot\partial_{k_y}\vec{\mathbb{E}}^*)\hat{\boldsymbol{y}}+(\vec{\mathbb{E}}\cdot\partial_{k_z}\vec{\mathbb{E}}^*)\hat{\boldsymbol{z}}\big]\times\boldsymbol{k}+(\vec{\mathbb{E}}^*\times\vec{\mathbb{E}})\}\mathrm{d}k_x\mathrm{d}k_y\mathrm{d}k_z.\qquad\text{(B9)}$$

In the above expression, the first term of the integrand, which is perpendicular to $\boldsymbol{k}$, represents the orbital as well as extrinsic components of the packet's angular momentum, whereas the second term, which is aligned with $\boldsymbol{k}$, corresponds to the spin component of the angular momentum.

**Digression 1**: To find the projection onto $\hat{\boldsymbol{k}}$ of the first term in the integrand of Eq.(B5) we need the following vector identity:

$$\boldsymbol{A}\cdot\boldsymbol{\nabla}(\boldsymbol{B}\cdot\boldsymbol{C}) = A_x\partial_x(\boldsymbol{B}\cdot\boldsymbol{C}) + A_y\partial_y(\boldsymbol{B}\cdot\boldsymbol{C}) + A_z\partial_z(\boldsymbol{B}\cdot\boldsymbol{C})$$
$$= A_x[(\partial_x\boldsymbol{B})\cdot\boldsymbol{C} + (\partial_x\boldsymbol{C})\cdot\boldsymbol{B}] + A_y[(\partial_y\boldsymbol{B})\cdot\boldsymbol{C} + (\partial_y\boldsymbol{C})\cdot\boldsymbol{B}] + A_z[(\partial_z\boldsymbol{B})\cdot\boldsymbol{C} + (\partial_z\boldsymbol{C})\cdot\boldsymbol{B}]$$
$$= [(\boldsymbol{A}\cdot\boldsymbol{\nabla})\boldsymbol{B}]\cdot\boldsymbol{C} + [(\boldsymbol{A}\cdot\boldsymbol{\nabla})\boldsymbol{C}]\cdot\boldsymbol{B}. \quad \text{(B10)}$$

If we now set $\boldsymbol{A}(\boldsymbol{k}) = \hat{\boldsymbol{k}}\times\vec{\mathbb{E}}(\boldsymbol{k})$, $\boldsymbol{B}(\boldsymbol{k}) = \vec{\mathbb{E}}^*(\boldsymbol{k})$, and $\boldsymbol{C}(\boldsymbol{k}) = \hat{\boldsymbol{k}}$, we will have $\boldsymbol{B}\cdot\boldsymbol{C} = 0$. Also, $\partial_{k_x}\hat{\boldsymbol{k}} = -(c/\omega)^3 k_x\boldsymbol{k} + (c/\omega)\hat{\boldsymbol{x}}$, and similar expressions exist for $\partial_{k_y}\hat{\boldsymbol{k}}$ and $\partial_{k_z}\hat{\boldsymbol{k}}$. Consequently,

$$[\{[\hat{\boldsymbol{k}}\times\vec{\mathbb{E}}(\boldsymbol{k})]\cdot\boldsymbol{\nabla}_{\boldsymbol{k}}\}\vec{\mathbb{E}}^*(\boldsymbol{k})]\cdot\hat{\boldsymbol{k}} = -(\{[\hat{\boldsymbol{k}}\times\vec{\mathbb{E}}(\boldsymbol{k})]\cdot\boldsymbol{\nabla}\}\hat{\boldsymbol{k}})\cdot\vec{\mathbb{E}}^*(\boldsymbol{k})$$
$$= [(c/\omega)^3\{\underbrace{[\hat{\boldsymbol{k}}\times\vec{\mathbb{E}}(\boldsymbol{k})]\cdot\boldsymbol{k}}_{0}\}\boldsymbol{k} - (c/\omega)\hat{\boldsymbol{k}}\times\vec{\mathbb{E}}(\boldsymbol{k})]\cdot\vec{\mathbb{E}}^*(\boldsymbol{k})$$
$$= -(c/\omega)[\hat{\boldsymbol{k}}\times\vec{\mathbb{E}}(\boldsymbol{k})]\cdot\vec{\mathbb{E}}^*(\boldsymbol{k}) = -(c/\omega)[\vec{\mathbb{E}}(\boldsymbol{k})\times\vec{\mathbb{E}}^*(\boldsymbol{k})]\cdot\hat{\boldsymbol{k}}$$

$\boldsymbol{k}\cdot\vec{\mathbb{E}} = k_x\mathbb{E}_x + k_y\mathbb{E}_y + k_z\mathbb{E}_z = 0$ →

$$= -(c/\omega)[(\mathbb{E}_y\mathbb{E}_z^* - \mathbb{E}_z\mathbb{E}_y^*)\hat{k}_x + (\mathbb{E}_z\mathbb{E}_x^* - \mathbb{E}_x\mathbb{E}_z^*)\hat{k}_y + (\mathbb{E}_x\mathbb{E}_y^* - \mathbb{E}_y\mathbb{E}_x^*)\hat{k}_z]$$
$$= -(c/\omega)(\mathbb{E}_x\mathbb{E}_y^* - \mathbb{E}_y\mathbb{E}_x^*)(k_x\hat{k}_x + k_y\hat{k}_y + k_z\hat{k}_z)/k_z$$
$$= -(c/\omega)(\mathbb{E}_x\mathbb{E}_y^* - \mathbb{E}_y\mathbb{E}_x^*)(\boldsymbol{k}\cdot\hat{\boldsymbol{k}})/k_z = -(\mathbb{E}_x\mathbb{E}_y^* - \mathbb{E}_y\mathbb{E}_x^*)/k_z. \quad \text{(B11)}$$

We have thus arrived at a simplified expression for the spin angular momentum, also found in Eq.(B7).

**Digression 2**: To find the component of the integrand that is perpendicular to $\hat{\boldsymbol{k}}$, we use the identity:

$$(\boldsymbol{A}\cdot\boldsymbol{\nabla})(\boldsymbol{B}\times\boldsymbol{C}) = A_x\partial_x(\boldsymbol{B}\times\boldsymbol{C}) + A_y\partial_y(\boldsymbol{B}\times\boldsymbol{C}) + A_z\partial_z(\boldsymbol{B}\times\boldsymbol{C})$$
$$= A_x[(\partial_x\boldsymbol{B})\times\boldsymbol{C} - (\partial_x\boldsymbol{C})\times\boldsymbol{B}] + A_y[(\partial_y\boldsymbol{B})\times\boldsymbol{C} - (\partial_y\boldsymbol{C})\times\boldsymbol{B}] + A_z[(\partial_z\boldsymbol{B})\times\boldsymbol{C} - (\partial_z\boldsymbol{C})\times\boldsymbol{B}]$$
$$= [(\boldsymbol{A}\cdot\boldsymbol{\nabla})\boldsymbol{B}]\times\boldsymbol{C} - [(\boldsymbol{A}\cdot\boldsymbol{\nabla})\boldsymbol{C}]\times\boldsymbol{B}. \quad \text{(B12)}$$

$$(\{[\hat{\boldsymbol{k}}\times\vec{\mathbb{E}}(\boldsymbol{k})]\cdot\boldsymbol{\nabla}_{\boldsymbol{k}}\}\vec{\mathbb{E}}^*(\boldsymbol{k}))\times\hat{\boldsymbol{k}} = \{[\hat{\boldsymbol{k}}\times\vec{\mathbb{E}}(\boldsymbol{k})]\cdot\boldsymbol{\nabla}_{\boldsymbol{k}}\}[\vec{\mathbb{E}}^*(\boldsymbol{k})\times\hat{\boldsymbol{k}}] + (\{[\hat{\boldsymbol{k}}\times\vec{\mathbb{E}}(\boldsymbol{k})]\cdot\boldsymbol{\nabla}_{\boldsymbol{k}}\}\hat{\boldsymbol{k}})\times\vec{\mathbb{E}}^*(\boldsymbol{k})$$
$$= (\hat{k}_y\mathbb{E}_z - \hat{k}_z\mathbb{E}_y)(\partial_{k_x}\vec{\mathbb{E}}^*\times\hat{\boldsymbol{k}} + \vec{\mathbb{E}}^*\times\partial_{k_x}\hat{\boldsymbol{k}})$$
$$+(\hat{k}_z\mathbb{E}_x - \hat{k}_x\mathbb{E}_z)(\partial_{k_y}\vec{\mathbb{E}}^*\times\hat{\boldsymbol{k}} + \vec{\mathbb{E}}^*\times\partial_{k_y}\hat{\boldsymbol{k}})$$
$$+(\hat{k}_x\mathbb{E}_y - \hat{k}_y\mathbb{E}_x)(\partial_{k_z}\vec{\mathbb{E}}^*\times\hat{\boldsymbol{k}} + \vec{\mathbb{E}}^*\times\partial_{k_z}\hat{\boldsymbol{k}})$$
$$+(c/\omega)[\hat{\boldsymbol{k}}\times\vec{\mathbb{E}}(\boldsymbol{k})]\times\vec{\mathbb{E}}^*(\boldsymbol{k})$$
$$= (\hat{k}_y\mathbb{E}_z - \hat{k}_z\mathbb{E}_y)[\partial_{k_x}\vec{\mathbb{E}}^*\times\hat{\boldsymbol{k}} + \cancel{(c/\omega)\hat{k}_x\hat{\boldsymbol{k}}\times\vec{\mathbb{E}}^*} + \cancel{(c/\omega)\vec{\mathbb{E}}^*\times\hat{\boldsymbol{x}}}]$$
$$+(\hat{k}_z\mathbb{E}_x - \hat{k}_x\mathbb{E}_z)[\partial_{k_y}\vec{\mathbb{E}}^*\times\hat{\boldsymbol{k}} + \cancel{(c/\omega)\hat{k}_y\hat{\boldsymbol{k}}\times\vec{\mathbb{E}}^*} + \cancel{(c/\omega)\vec{\mathbb{E}}^*\times\hat{\boldsymbol{y}}}]$$
$$+(\hat{k}_x\mathbb{E}_y - \hat{k}_y\mathbb{E}_x)[\partial_{k_z}\vec{\mathbb{E}}^*\times\hat{\boldsymbol{k}} + \cancel{(c/\omega)\hat{k}_z\hat{\boldsymbol{k}}\times\vec{\mathbb{E}}^*} + \cancel{(c/\omega)\vec{\mathbb{E}}^*\times\hat{\boldsymbol{z}}}]$$
$$\cancel{-(c/\omega)(\vec{\mathbb{E}}\cdot\vec{\mathbb{E}}^*)}\hat{\boldsymbol{k}}$$
$$= [(\hat{k}_y\mathbb{E}_z - \hat{k}_z\mathbb{E}_y)\partial_{k_x}\vec{\mathbb{E}}^* + (\hat{k}_z\mathbb{E}_x - \hat{k}_x\mathbb{E}_z)\partial_{k_y}\vec{\mathbb{E}}^* + (\hat{k}_x\mathbb{E}_y - \hat{k}_y\mathbb{E}_x)\partial_{k_z}\vec{\mathbb{E}}^*]\times\hat{\boldsymbol{k}}. \quad \text{(B13)}$$

We are now back where we had started. This approach, therefore, does *not* yield the desired result.

## Appendix C
## Radiation from a spherical dipole

The radiated $E$ and $H$ fields of a spherical dipole $\boldsymbol{P}(t)$, which has been active during the time interval $0 \le t \le T$, were evaluated at $t_0 > T$ and given in Eqs.(19a) and (19b). To compute the total radiated energy associated with these fields we must integrate the corresponding energy densities over the entire $xyz$ space. Upon invoking the identity $\int_{-\infty}^{\infty} \exp[\mathrm{i}(\boldsymbol{k}+\boldsymbol{k}')\cdot\boldsymbol{r}]\,\mathrm{d}\boldsymbol{r} = (2\pi)^3\delta(\boldsymbol{k}+\boldsymbol{k}')$, we find

$$\int_{-\infty}^{\infty} \tfrac{1}{2}\varepsilon_0 \boldsymbol{E}(\boldsymbol{r},t_0)\cdot\boldsymbol{E}(\boldsymbol{r},t_0)\mathrm{d}\boldsymbol{r} = \frac{c^2}{\pi\varepsilon_0}\int_{-\infty}^{\infty}\{[\sin(kR) - kR\cos(kR)]/k^4\}^2$$
$$\times\left\{\{k^2\vec{\mathbb{P}}_R(ck) - [\vec{\mathbb{P}}_R(ck)\cdot\boldsymbol{k}]\boldsymbol{k}\}\sin(ckt_0) - \{k^2\vec{\mathbb{P}}_I(ck) - [\vec{\mathbb{P}}_I(ck)\cdot\boldsymbol{k}]\boldsymbol{k}\}\cos(ckt_0)\right\}^2\mathrm{d}\boldsymbol{k}. \quad \text{(C1)}$$

$$\int_{-\infty}^{\infty} \tfrac{1}{2}\mu_0 \boldsymbol{H}(\boldsymbol{r},t_0)\cdot\boldsymbol{H}(\boldsymbol{r},t_0)\mathrm{d}\boldsymbol{r} = \frac{c^2}{\pi\varepsilon_0}\int_{-\infty}^{\infty}\{[\sin(kR) - kR\cos(kR)]/k^3\}^2$$
$$\times[\boldsymbol{k}\times\vec{\mathbb{P}}_R(ck)\cos(ckt_0) + \boldsymbol{k}\times\vec{\mathbb{P}}_I(ck)\sin(ckt_0)]^2\mathrm{d}\boldsymbol{k}. \quad \text{(C2)}$$

square in the sense of dot-product

Adding Eqs.(C1) and (C2), we find that the time-dependent part of the total radiated energy vanishes (provided, of course, that $t_0 > T$), as confirmed below:

$$\{k^2(\boldsymbol{k}\times\mathbb{P}_R)\cdot(\boldsymbol{k}\times\mathbb{P}_R) - [k^2\mathbb{P}_R - (\mathbb{P}_R\cdot\boldsymbol{k})\boldsymbol{k}]\cdot[k^2\mathbb{P}_R - (\mathbb{P}_R\cdot\boldsymbol{k})\boldsymbol{k}]\}\cos(2ckt_0)$$
$$-\{k^2(\boldsymbol{k}\times\mathbb{P}_I)\cdot(\boldsymbol{k}\times\mathbb{P}_I) - [k^2\mathbb{P}_I - (\mathbb{P}_I\cdot\boldsymbol{k})\boldsymbol{k}]\cdot[k^2\mathbb{P}_I - (\mathbb{P}_I\cdot\boldsymbol{k})\boldsymbol{k}]\}\cos(2ckt_0)$$
$$+\{k^2(\boldsymbol{k}\times\mathbb{P}_R)\cdot(\boldsymbol{k}\times\mathbb{P}_I) - [k^2\mathbb{P}_R - (\mathbb{P}_R\cdot\boldsymbol{k})\boldsymbol{k}]\cdot[k^2\mathbb{P}_I - (\mathbb{P}_I\cdot\boldsymbol{k})\boldsymbol{k}]\}\sin(2ckt_0)$$
$$= k^2\big[(\boldsymbol{k}\times\mathbb{P}_R)\cdot(\boldsymbol{k}\times\mathbb{P}_R) + (\mathbb{P}_R\cdot\boldsymbol{k})^{\boldsymbol{2}} - k^2\mathbb{P}_R\cdot\mathbb{P}_R\big]\cos(2ckt_0)$$
$$-k^2\big[(\boldsymbol{k}\times\mathbb{P}_I)\cdot(\boldsymbol{k}\times\mathbb{P}_I) + (\mathbb{P}_I\cdot\boldsymbol{k})^{\boldsymbol{2}} - k^2\mathbb{P}_I\cdot\mathbb{P}_I\big]\cos(2ckt_0)$$
$$+k^2\big[(\boldsymbol{k}\times\vec{\mathbb{P}}_R)\cdot(\boldsymbol{k}\times\vec{\mathbb{P}}_I) + (\vec{\mathbb{P}}_R\cdot\boldsymbol{k})(\vec{\mathbb{P}}_I\cdot\boldsymbol{k}) - k^2\vec{\mathbb{P}}_R\cdot\vec{\mathbb{P}}_I\big]\sin(2ckt_0) = 0. \quad \text{(C3)}$$

The remaining terms combine to yield

$$\tfrac{1}{2}[k^2\mathbb{P}_R - (\mathbb{P}_R\cdot\boldsymbol{k})\boldsymbol{k}]\cdot[k^2\mathbb{P}_R - (\mathbb{P}_R\cdot\boldsymbol{k})\boldsymbol{k}] + \tfrac{1}{2}[k^2\mathbb{P}_I - (\mathbb{P}_I\cdot\boldsymbol{k})\boldsymbol{k}]\cdot[k^2\mathbb{P}_I - (\mathbb{P}_I\cdot\boldsymbol{k})\boldsymbol{k}]$$
$$+\tfrac{1}{2}k^2[(\boldsymbol{k}\times\mathbb{P}_R)\cdot(\boldsymbol{k}\times\mathbb{P}_R) + (\boldsymbol{k}\times\mathbb{P}_I)\cdot(\boldsymbol{k}\times\mathbb{P}_I)]$$
$$= k^4\big[\mathbb{P}_R\cdot\mathbb{P}_R + \mathbb{P}_I\cdot\mathbb{P}_I - (\mathbb{P}_R\cdot\hat{\boldsymbol{k}})^2 - (\mathbb{P}_I\cdot\hat{\boldsymbol{k}})^2\big] = k^2[(\mathbb{P}_R\times\boldsymbol{k})^2 + (\mathbb{P}_I\times\boldsymbol{k})^2]. \quad \text{(C4)}$$

The total radiated energy of the dipole is thus given by

$$\mathcal{E}(t_0) = \frac{R^6}{\pi\varepsilon_0}\iiint_{-\infty}^{\infty}\left[\frac{\sin(kR) - kR\cos(kR)}{(kR)^3}\right]^2\{[\vec{\mathbb{P}}_R(ck)\times c\boldsymbol{k}]^2 + [\vec{\mathbb{P}}_I(ck)\times c\boldsymbol{k}]^2\}\mathrm{d}\boldsymbol{k}$$
$$= \frac{R^6}{\pi\varepsilon_0}\iiint_{-\infty}^{\infty}\left[\frac{\sin(kR) - kR\cos(kR)}{(kR)^3}\right]^2[(\vec{\mathbb{P}}_R + \mathrm{i}\vec{\mathbb{P}}_I)\times c\boldsymbol{k}]\cdot[(\vec{\mathbb{P}}_R - \mathrm{i}\vec{\mathbb{P}}_I)\times c\boldsymbol{k}]\mathrm{d}\boldsymbol{k}$$
$$= \frac{R^6c^2}{\pi\varepsilon_0}\iiint_{-\infty}^{\infty}\left[\frac{\sin(kR) - kR\cos(kR)}{(kR)^3}\right]^2\big[k^2\vec{\mathbb{P}}\cdot\vec{\mathbb{P}}^* - (\boldsymbol{k}\cdot\vec{\mathbb{P}})(\boldsymbol{k}\cdot\vec{\mathbb{P}}^*)\big]\mathrm{d}\boldsymbol{k}. \quad \text{(C5)}$$

In spherical coordinates in $k$-space, where $k_x = k\sin\theta\cos\varphi$, $k_y = k\sin\theta\sin\varphi$, and $k_z = k\cos\theta$, we may write

$$(\boldsymbol{k}\cdot\vec{\mathbb{P}})(\boldsymbol{k}\cdot\vec{\mathbb{P}}^*) = k_x^2\mathbb{P}_x\mathbb{P}_x^* + k_y^2\mathbb{P}_y\mathbb{P}_y^* + k_z^2\mathbb{P}_z\mathbb{P}_z^* + k_xk_y(\mathbb{P}_x\mathbb{P}_y^* + \mathbb{P}_y\mathbb{P}_x^*)$$
$$+k_xk_z(\mathbb{P}_x\mathbb{P}_z^* + \mathbb{P}_z\mathbb{P}_x^*) + k_yk_z(\mathbb{P}_y\mathbb{P}_z^* + \mathbb{P}_z\mathbb{P}_y^*). \quad \text{(C6)}$$

Considering that $\mathbb{P}(ck)$ is independent of $\theta$ and $\varphi$, integration over the $(\theta,\varphi)$ coordinates yields

$$\int_{\theta=0}^{\pi}\int_{\varphi=0}^{2\pi}(\boldsymbol{k}\cdot\vec{\mathbb{P}})(\boldsymbol{k}\cdot\vec{\mathbb{P}}^*)k^2\sin\theta\,\mathrm{d}\theta\mathrm{d}\varphi = k^4\mathbb{P}_x\mathbb{P}_x^*\int_{\theta=0}^{\pi}\int_{\varphi=0}^{2\pi}\sin^3\theta\cos^2\varphi\,\mathrm{d}\theta\mathrm{d}\varphi \quad \leftarrow \boxed{\textstyle\int_0^\pi \sin^3\theta\,\mathrm{d}\theta = 4/3}$$

$$+k^4\mathbb{P}_y\mathbb{P}_y^*\int_{\theta=0}^{\pi}\int_{\varphi=0}^{2\pi}\sin^3\theta\sin^2\varphi\,\mathrm{d}\theta\mathrm{d}\varphi + k^4\mathbb{P}_z\mathbb{P}_z^*\int_{\theta=0}^{\pi}\int_{\varphi=0}^{2\pi}\sin\theta\cos^2\theta\,\mathrm{d}\theta\mathrm{d}\varphi$$

$$\uparrow \boxed{\textstyle\int_0^\pi \sin\theta\cos^2\theta\mathrm{d}\theta = 2/3}$$

$$= (4\pi/3)k^4(\mathbb{P}_x\mathbb{P}_x^* + \mathbb{P}_y\mathbb{P}_y^* + \mathbb{P}_z\mathbb{P}_z^*) = (4\pi/3)k^4\vec{\mathbb{P}}(ck)\cdot\vec{\mathbb{P}}^*(ck). \tag{C7}$$

Equation (C5) is thus simplified, as follows:

$$\mathcal{E}(t_0) = \frac{8R^6c^2}{3\varepsilon_0}\int_{k=0}^{\infty}\left[\frac{\sin(kR)-kR\cos(kR)}{(kR)^3}\right]^2 k^4\vec{\mathbb{P}}(ck)\cdot\vec{\mathbb{P}}^*(ck)\mathrm{d}k$$

$$\boxed{kR \ll 1} \rightarrow \ \cong \frac{8R^6c^2}{27\varepsilon_0}\int_0^{\infty}k^4\vec{\mathbb{P}}(ck)\cdot\vec{\mathbb{P}}^*(ck)\mathrm{d}k = \frac{8\mu_0R^6}{27c}\int_0^{\infty}\omega^4\vec{\mathbb{P}}(\omega)\cdot\vec{\mathbb{P}}^*(\omega)\mathrm{d}\omega. \tag{C8}$$

---

**Example 1**. Assuming that $\vec{\mathbb{P}}_R(\omega) = \mathbb{P}_R(\omega)\hat{\boldsymbol{x}}$ and $\vec{\mathbb{P}}_I(\omega) = \mathbb{P}_I(\omega)\hat{\boldsymbol{y}}$, the total radiated energy at $t = t_0$ will be given by Eq.(C5), as follows:

$$\mathcal{E}(t_0) = \frac{R^6c^2}{\pi\varepsilon_0}\int_{-\infty}^{\infty}\left[\frac{\sin(kR)-kR\cos(kR)}{(kR)^3}\right]^2\left[\mathbb{P}_R^2(ck)(k_y^2+k_z^2) + \mathbb{P}_I^2(ck)(k_x^2+k_z^2)\right]\mathrm{d}\boldsymbol{k}$$

$$= \frac{R^6c^2}{\pi\varepsilon_0}\int_{k=0}^{\infty}\int_{\theta=0}^{\pi}\int_{\varphi=0}^{2\pi}\left[\frac{\sin(kR)-kR\cos(kR)}{(kR)^3}\right]^2 k^4\sin\theta$$

$$\times\left[\mathbb{P}_R^2(ck)(\sin^2\theta\sin^2\varphi+\cos^2\theta) + \mathbb{P}_I^2(ck)(\sin^2\theta\cos^2\varphi+\cos^2\theta)\right]\mathrm{d}k\mathrm{d}\theta\mathrm{d}\varphi$$

$$= \frac{R^6c^2}{\varepsilon_0}\int_{k=0}^{\infty}\left[\frac{\sin(kR)-kR\cos(kR)}{(kR)^3}\right]^2 k^4[\mathbb{P}_R^2(ck)+\mathbb{P}_I^2(ck)]\int_{\theta=0}^{\pi}\sin\theta\,(1+\cos^2\theta)\mathrm{d}\theta\mathrm{d}k$$

$$= \frac{8\mu_0R^6}{3c}\int_0^{\infty}\left[\frac{\sin(R\omega/c)-(R\omega/c)\cos(R\omega/c)}{(R\omega/c)^3}\right]^2\omega^4[\mathbb{P}_R^2(\omega)+\mathbb{P}_I^2(\omega)]\mathrm{d}\omega. \tag{C9}$$

---

**Example 2**. Let $\boldsymbol{P}(t) = P_0\cos(\omega_0 t)\exp[-(t/T)^2]\,\hat{\boldsymbol{z}}$. we will have

$$\vec{\mathbb{P}}(\omega) = \tfrac{1}{2}P_0\hat{\boldsymbol{z}}\int_{-\infty}^{\infty}\exp[-(t/T)^2]\{\exp[\mathrm{i}(\omega+\omega_0)t] + \exp[\mathrm{i}(\omega-\omega_0)t]\}\mathrm{d}t$$

$$= \tfrac{1}{2}P_0\exp[-\tfrac{1}{4}T^2(\omega+\omega_0)^2]\hat{\boldsymbol{z}}\int_{-\infty}^{\infty}\exp\{-[t-\tfrac{1}{2}\mathrm{i}T^2(\omega+\omega_0)]^2/T^2\}\mathrm{d}t$$

$$+\tfrac{1}{2}P_0\exp[-\tfrac{1}{4}T^2(\omega-\omega_0)^2]\hat{\boldsymbol{z}}\int_{-\infty}^{\infty}\exp\{-[t-\tfrac{1}{2}\mathrm{i}T^2(\omega-\omega_0)]^2/T^2\}\mathrm{d}t$$

$$= \tfrac{1}{2}\sqrt{\pi}P_0T\{\exp[-\tfrac{1}{4}T^2(\omega+\omega_0)^2] + \exp[-\tfrac{1}{4}T^2(\omega-\omega_0)^2]\}\hat{\boldsymbol{z}}. \tag{C10}$$

If $R$ is sufficiently small, we will have $[\sin(kR) - kR\cos(kR)]/(kR)^3 \cong \tfrac{1}{3}$, in which case Eq.(C5) yields

$$\mathcal{E}(t_0) \cong \frac{R^6}{9\pi\varepsilon_0}\int_{-\infty}^{\infty}\left\{\tfrac{1}{2}\sqrt{\pi}P_0T\exp[-\tfrac{1}{4}T^2(ck-\omega_0)^2]\hat{\boldsymbol{z}}\times c\boldsymbol{k}\right\}^2\mathrm{d}\boldsymbol{k} \quad \leftarrow \boxed{\text{For sufficiently large } \omega_0,\ \exp[-\tfrac{1}{4}T^2(ck+\omega_0)^2] \text{ is small and may be ignored.}}$$

$$= \frac{c^2P_0^2T^2R^6}{36\varepsilon_0}\int_{k=0}^{\infty}\int_{\theta=0}^{\pi}2\pi\exp[-\tfrac{1}{2}T^2(ck-\omega_0)^2]k^4\sin^3\theta\,\mathrm{d}k\mathrm{d}\theta \quad \leftarrow \boxed{\textstyle\int_0^\pi \sin^3\theta\,d\theta = 4/3}$$

$$\cong \frac{2\pi P_0^2TR^6\omega_0^4}{27\varepsilon_0c^3}\int_{-\infty}^{\infty}\exp(-\tfrac{1}{2}x^2)\,\mathrm{d}x = \frac{\mu_0p_0^2\omega_0^4}{12\pi c}\sqrt{\pi/2}\,T. \tag{C11}$$

The dipolar intensity profile $\exp[-2(t/T)^2]$ has width $= \sqrt{\pi/2}\,T$. The remaining coefficient on the right-hand-side of Eq.(C11) is the rate of radiation of EM energy from the point-dipole $\boldsymbol{p}_0 = (4\pi R^3/3)P_0\cos(\omega_0 t)\,\hat{\boldsymbol{z}}$.

---

**Rate of energy extraction from the dipole**. The energy extraction rate (per unit volume per unit time) is given by $\boldsymbol{E}\cdot\partial\boldsymbol{P}/\partial t$. We thus write

$$\int_{-\infty}^{\infty}\int_{t=0}^{T}\boldsymbol{E}\cdot(\partial\boldsymbol{P}/\partial t)\mathrm{d}\boldsymbol{r}\mathrm{d}t=\frac{1}{2\pi^3\varepsilon_0}\int_{-\infty}^{\infty}\left[\frac{\sin(kR)-kR\cos(kR)}{k^3}\right]^2\left(\frac{c}{k}\right)\left\{\frac{k/c}{k^2-(\omega/c)^2}+½\mathrm{i}\pi[\delta(\omega-ck)-\delta(\omega+ck)]\right\}$$

$$\times\left\{(\omega/c)^2\vec{\mathbb{P}}(\omega)-[\vec{\mathbb{P}}(\omega)\cdot\boldsymbol{k}]\boldsymbol{k}\right\}\cdot[-\mathrm{i}\omega'\vec{\mathbb{P}}(\omega')]\left\{\int_{-\infty}^{\infty}\exp[-\mathrm{i}(\omega+\omega')t]\,\mathrm{d}t\right\}\mathrm{d}\omega\mathrm{d}\omega'\mathrm{d}\boldsymbol{k}$$

$$=\frac{1}{\pi^2\varepsilon_0}\int_{-\infty}^{\infty}\left[\frac{\sin(kR)-kR\cos(kR)}{k^3}\right]^2$$

$$\times\int_{-\infty}^{\infty}\left(\frac{c}{k}\right)\left\{\frac{k/c}{k^2-(\omega/c)^2}+½\mathrm{i}\pi[\delta(\omega-ck)-\delta(\omega+ck)]\right\}\left\{(\omega/c)^2\vec{\mathbb{P}}(\omega)-[\vec{\mathbb{P}}(\omega)\cdot\boldsymbol{k}]\boldsymbol{k}\right\}\cdot[\mathrm{i}\omega\vec{\mathbb{P}}^*(\omega)]\mathrm{d}\omega\mathrm{d}\boldsymbol{k}$$

$$=-\frac{c^2}{2\pi\varepsilon_0}\int_{-\infty}^{\infty}\left[\frac{\sin(kR)-kR\cos(kR)}{k^3}\right]^2\int_{-\infty}^{\infty}[\delta(\omega-ck)+\delta(\omega+ck)]\left\{(\omega/c)^2\vec{\mathbb{P}}(\omega)-[\vec{\mathbb{P}}(\omega)\cdot\boldsymbol{k}]\boldsymbol{k}\right\}\cdot\vec{\mathbb{P}}^*(\omega)\mathrm{d}\omega\mathrm{d}\boldsymbol{k}$$

$$=-\frac{c^2}{\pi\varepsilon_0}\int_{-\infty}^{\infty}\left[\frac{\sin(kR)-kR\cos(kR)}{k^3}\right]^2\left\{k^2\vec{\mathbb{P}}(ck)\cdot\vec{\mathbb{P}}^*(ck)-[\vec{\mathbb{P}}(ck)\cdot\boldsymbol{k}][\vec{\mathbb{P}}^*(ck)\cdot\boldsymbol{k}]\right\}\mathrm{d}\boldsymbol{k}$$

$$=-\frac{c^2}{\pi\varepsilon_0}\int_{-\infty}^{\infty}\left[\frac{\sin(kR)-kR\cos(kR)}{k^3}\right]^2\left\{k^2\vec{\mathbb{P}}_R(ck)\cdot\vec{\mathbb{P}}_R(ck)+k^2\vec{\mathbb{P}}_I(ck)\cdot\vec{\mathbb{P}}_I(ck)-[\vec{\mathbb{P}}_R(ck)\cdot\boldsymbol{k}]^2-[\vec{\mathbb{P}}_I(ck)\cdot\boldsymbol{k}]^2\right\}\mathrm{d}\boldsymbol{k}$$

$$=-\frac{R^6}{\pi\varepsilon_0}\int_{-\infty}^{\infty}\left[\frac{\sin(kR)-kR\cos(kR)}{(kR)^3}\right]^2\left\{[\vec{\mathbb{P}}_R(ck)\times c\boldsymbol{k}]^2+[\vec{\mathbb{P}}_I(ck)\times c\boldsymbol{k}]^2\right\}\mathrm{d}\boldsymbol{k}.\qquad\text{(C12)}$$

The above equation is identical to Eq.(C5), except for an overall minus sign, which indicates that the energy in Eq.(C12) is *extracted* from the dipole.

**Cross-terms between the incident wave-packet and the radiated field of the spherical dipole**. The total $E$-field energy at $t=t_0>T$ is an integral over the entire $xyz$ space of the sum of the incident $E$-field, $\boldsymbol{E}^{(\mathrm{i})}(\boldsymbol{r},t_0)$, and the radiated $E$-field, $\boldsymbol{E}^{(\mathrm{r})}(\boldsymbol{r},t_0)$, that is,

$$\mathcal{E}_E(t_0)=½\varepsilon_0\int[\boldsymbol{E}^{(\mathrm{i})}(\boldsymbol{r},t_0)+\boldsymbol{E}^{(\mathrm{r})}(\boldsymbol{r},t_0)]\cdot[\boldsymbol{E}^{(\mathrm{i})}(\boldsymbol{r},t_0)+\boldsymbol{E}^{(\mathrm{r})}(\boldsymbol{r},t_0)]\mathrm{d}\boldsymbol{r}$$

$$=½\varepsilon_0\int\boldsymbol{E}^{(\mathrm{i})}(\boldsymbol{r},t_0)\cdot\boldsymbol{E}^{(\mathrm{i})}(\boldsymbol{r},t_0)\mathrm{d}\boldsymbol{r}+½\varepsilon_0\int\boldsymbol{E}^{(\mathrm{r})}(\boldsymbol{r},t_0)\cdot\boldsymbol{E}^{(\mathrm{r})}(\boldsymbol{r},t_0)\mathrm{d}\boldsymbol{r}+\varepsilon_0\int\boldsymbol{E}^{(\mathrm{i})}(\boldsymbol{r},t_0)\cdot\boldsymbol{E}^{(\mathrm{r})}(\boldsymbol{r},t_0)\mathrm{d}\boldsymbol{r}.\qquad\text{(C13)}$$

Considering that the incident wave-packet has uniform cross-section in the $xy$-plane, it may be expanded as a superposition of plane-waves of differing $k_z$ or differing $\omega$, where $\omega=ck_z$, namely,

$$\boldsymbol{E}^{(\mathrm{i})}(\boldsymbol{r},t)=\frac{1}{2\pi}\int_{-\infty}^{\infty}\vec{\mathbb{E}}^{(\mathrm{i})}(k_z)\exp[\mathrm{i}k_z(z-ct)]\,\mathrm{d}k_z=\frac{1}{2\pi}\int_{-\infty}^{\infty}\boldsymbol{\mathcal{E}}^{(\mathrm{i})}(\omega)\exp[\mathrm{i}(\omega/c)(z-ct)]\,\mathrm{d}\omega.\qquad\text{(C14)}$$

In the above equation, $\vec{\mathbb{E}}^{(\mathrm{i})}(k_z)=c\boldsymbol{\mathcal{E}}^{(\mathrm{i})}(\omega)=c\boldsymbol{\mathcal{E}}^{(\mathrm{i})}(ck_z)$. In the International System of Units (SI), the units of $\boldsymbol{E}(\boldsymbol{r},t)$ are volt/m, while those of $\vec{\mathbb{E}}(k_z)$ and $\boldsymbol{\mathcal{E}}(\omega)$ are volt and volt·sec/m, respectively. The polarization $\boldsymbol{P}(t)$ of the spherical dipole is now seen to be related to the incident $E$-field as follows:

$$\boldsymbol{P}(t)=\frac{\varepsilon_0}{2\pi}\int_{-\infty}^{\infty}\chi(\omega)\boldsymbol{\mathcal{E}}^{(\mathrm{i})}(\omega)\exp(-\mathrm{i}\omega t)\,\mathrm{d}\omega\;\rightarrow\;\vec{\mathbb{P}}(\omega)=\varepsilon_0\chi(\omega)\boldsymbol{\mathcal{E}}^{(\mathrm{i})}(\omega)=(\varepsilon_0/c)\chi(ck_z)\vec{\mathbb{E}}^{(\mathrm{i})}(k_z).\qquad\text{(C15)}$$

Here $\chi(\omega)=\chi'(\omega)+\mathrm{i}\chi''(\omega)$ is the susceptibility of the spherical volume of material to the incident $E$-field. The $E$-field energy of the cross-term integrated over all space is now obtained as follows:

$$\varepsilon_0\int\boldsymbol{E}^{(\mathrm{i})}\cdot\boldsymbol{E}^{(\mathrm{r})}\mathrm{d}\boldsymbol{r}=\frac{\varepsilon_0}{(2\pi)(4\pi^3)\varepsilon_0}\int\vec{\mathbb{E}}^{(\mathrm{i})}(k_z')\exp[\mathrm{i}k_z'(z-ct_0)]\cdot\vec{\mathbb{E}}^{(\mathrm{r})}(\boldsymbol{k},\omega)\exp[\mathrm{i}(\boldsymbol{k}\cdot\boldsymbol{r}-\omega t_0)]\,\mathrm{d}\boldsymbol{r}\mathrm{d}k_z'\mathrm{d}\boldsymbol{k}\mathrm{d}\omega$$

$$=\frac{1}{\pi}\int\frac{\vec{\mathbb{E}}^{(\mathrm{i})}(-k_z)\exp(\mathrm{i}ck_zt_0)\cdot\{(\omega/c)^2\vec{\mathbb{P}}(\omega)-[\vec{\mathbb{P}}(\omega)\cdot\boldsymbol{k}]\boldsymbol{k}\}[\sin(k_zR)-k_zR\cos(k_zR)]\exp(-\mathrm{i}\omega t_0)}{k_z^3[k_z^2-(\omega/c)^2]}\mathrm{d}\omega\mathrm{d}k_z$$

$$= \frac{2\pi c}{\pi} \int \frac{[\sin(k_z R) - k_z R \cos(k_z R)] \vec{\mathbb{E}}^{(\mathrm{i})}(-k_z) \exp(\mathrm{i} c k_z t_0)}{k_z^4}$$

$$\cdot \left[ \{ k_z^2 \vec{\mathbb{P}}_R(ck_z) - \cancel{[\vec{\mathbb{P}}_R(ck_z) \cdot \boldsymbol{k}]\boldsymbol{k}} \} \sin(ck_z t_0) - \{ k_z^2 \vec{\mathbb{P}}_I(ck_z) - \cancel{[\vec{\mathbb{P}}_I(ck_z) \cdot \boldsymbol{k}]\boldsymbol{k}} \} \cos(ck_z t_0) \right] \mathrm{d}k_z$$

$$= 2c \int \frac{[\sin(k_z R) - k_z R \cos(k_z R)] \exp(\mathrm{i} c k_z t_0)}{k_z^2}$$

$$\times \vec{\mathbb{E}}^{(\mathrm{i})*}(k_z) \cdot \left[ \vec{\mathbb{P}}_R(ck_z) \sin(ck_z t_0) - \vec{\mathbb{P}}_I(ck_z) \cos(ck_z t_0) \right] \mathrm{d}k_z$$

$$= c \int \left[ \frac{\sin(k_z R) - k_z R \cos(k_z R)}{k_z^2} \right] \vec{\mathbb{E}}^{(\mathrm{i})*}(k_z)$$

$$\cdot \left\{ \mathrm{i} \vec{\mathbb{P}}_R(ck_z)[1 - \exp(2\mathrm{i} c k_z t_0)] - \vec{\mathbb{P}}_I(ck_z)[1 + \exp(2\mathrm{i} c k_z t_0)] \right\} \mathrm{d}k_z$$

$$= \mathrm{i} c \int \left[ \frac{\sin(k_z R) - k_z R \cos(k_z R)}{k_z^2} \right] \vec{\mathbb{E}}^{(\mathrm{i})*}(k_z) \cdot \left[ \vec{\mathbb{P}}(ck_z) - \vec{\mathbb{P}}^*(ck_z) \exp(2\mathrm{i} c k_z t_0) \right] \mathrm{d}k_z$$

$$= \mathrm{i}\varepsilon_0 \int_{-\infty}^{\infty} \left[ \frac{\sin(k_z R) - k_z R \cos(k_z R)}{k_z^2} \right] \left[ \chi(ck_z) \vec{\mathbb{E}}^{(\mathrm{i})}(k_z) - \chi^*(ck_z) \vec{\mathbb{E}}^{(\mathrm{i})*}(k_z) \exp(\mathrm{i}2ck_z t_0) \right] \cdot \vec{\mathbb{E}}^{(\mathrm{i})*}(k_z) \mathrm{d}k_z. \quad \text{(C16)}$$

odd function of $k_z$ | Hermitian | Hermitian

Similarly, the $H$-field energy at $t = t_0 > T$, integrated over the entire $xyz$ space, is written as

$$\mathcal{E}_H(t) = ½\mu_0 \int [\boldsymbol{H}^{(\mathrm{i})}(\boldsymbol{r}, t) + \boldsymbol{H}^{(\mathrm{r})}(\boldsymbol{r}, t)] \cdot [\boldsymbol{H}^{(\mathrm{i})}(\boldsymbol{r}, t) + \boldsymbol{H}^{(\mathrm{r})}(\boldsymbol{r}, t)] \mathrm{d}\boldsymbol{r}$$

$$= ½\mu_0 \int \boldsymbol{H}^{(\mathrm{i})}(\boldsymbol{r}, t) \cdot \boldsymbol{H}^{(\mathrm{i})}(\boldsymbol{r}, t) \mathrm{d}\boldsymbol{r} + ½\mu_0 \int \boldsymbol{H}^{(\mathrm{r})}(\boldsymbol{r}, t) \cdot \boldsymbol{H}^{(\mathrm{r})}(\boldsymbol{r}, t) \mathrm{d}\boldsymbol{r} + \mu_0 \int \boldsymbol{H}^{(\mathrm{i})}(\boldsymbol{r}, t) \cdot \boldsymbol{H}^{(\mathrm{r})}(\boldsymbol{r}, t) \mathrm{d}\boldsymbol{r}. \quad \text{(C17)}$$

The energy associated with the cross-term between the incident and radiated $H$-fields is thus given by

$$\mu_0 \int \boldsymbol{H}^{(\mathrm{i})} \cdot \boldsymbol{H}^{(\mathrm{r})} \mathrm{d}\boldsymbol{r} = \frac{\mu_0}{(2\pi Z_0)(4\pi^3)} \int_{-\infty}^{\infty} [\hat{\boldsymbol{z}} \times \vec{\mathbb{E}}^{(\mathrm{i})}(k_z')] \exp[\mathrm{i}k_z'(z - ct_0)] \cdot \vec{\mathbb{H}}^{(\mathrm{r})}(\boldsymbol{k}, \omega) \exp[\mathrm{i}(\boldsymbol{k} \cdot \boldsymbol{r} - \omega t_0)] \, \mathrm{d}\boldsymbol{r} \mathrm{d}k_z' \mathrm{d}\boldsymbol{k} \mathrm{d}\omega$$

$$= \frac{1}{\pi c} \int_{-\infty}^{\infty} [\hat{\boldsymbol{z}} \times \vec{\mathbb{E}}^{(\mathrm{i})}(-k_z)] \exp(\mathrm{i}ck_z t_0) \cdot \frac{k_z \hat{\boldsymbol{z}} \times \omega \vec{\mathbb{P}}(\omega) [\sin(k_z R) - k_z R \cos(k_z R)]}{k_z^3 [k_z^2 - (\omega/c)^2]} \exp(-\mathrm{i}\omega t_0) \, \mathrm{d}\omega \mathrm{d}k_z$$

$$= \frac{2\pi \mathrm{i}}{\pi} \int_{-\infty}^{\infty} \left[ \frac{\sin(k_z R) - k_z R \cos(k_z R)}{k_z^3} \right] [\hat{\boldsymbol{z}} \times \vec{\mathbb{E}}^{(\mathrm{i})*}(k_z)] \exp(\mathrm{i}ck_z t_0)$$

$$\cdot \left\{ ck_z \hat{\boldsymbol{z}} \times [\cos(ck_z t_0) \vec{\mathbb{P}}_R(ck_z) + \sin(ck_z t_0) \vec{\mathbb{P}}_I(ck_z)] \right\} \mathrm{d}k_z$$

$$= \mathrm{i}2c \int_{-\infty}^{\infty} \left[ \frac{\sin(k_z R) - k_z R \cos(k_z R)}{k_z^2} \right] \exp(\mathrm{i}ck_z t_0)$$

$$\vec{\mathbb{E}}^{(\mathrm{i})*}(k_z) \cdot [\cos(ck_z t_0) \vec{\mathbb{P}}_R(ck_z) + \sin(ck_z t_0) \vec{\mathbb{P}}_I(ck_z)] \mathrm{d}k_z$$

$$= \mathrm{i}c \int_{-\infty}^{\infty} \left[ \frac{\sin(k_z R) - k_z R \cos(k_z R)}{k_z^2} \right] \vec{\mathbb{E}}^{(\mathrm{i})*}(k_z)$$

$$\cdot \{ \vec{\mathbb{P}}_R(ck_z)[1 + \exp(2\mathrm{i}ck_z t_0)] + \mathrm{i}\vec{\mathbb{P}}_I(ck_z)[1 - \exp(2\mathrm{i}ck_z t_0)] \} \mathrm{d}k_z$$

$$= \mathrm{i}\varepsilon_0 \int_{-\infty}^{\infty} \left[ \frac{\sin(k_z R) - k_z R \cos(k_z R)}{k_z^2} \right] \left[ \chi(ck_z) \vec{\mathbb{E}}^{(\mathrm{i})}(k_z) + \chi^*(ck_z) \vec{\mathbb{E}}^{(\mathrm{i})*}(k_z) \exp(\mathrm{i}2ck_z t_0) \right] \cdot \vec{\mathbb{E}}^{(\mathrm{i})*}(k_z) \mathrm{d}k_z. \quad \text{(C18)}$$

Adding up the total cross-term energies of the $E$ and $H$ fields given by Eqs.(C16) and (C18), we now find

$$\mathcal{E}_{\text{total}}^{(\text{cross})}(t_0) = \mathcal{E}_E + \mathcal{E}_H = -2\varepsilon_0 R^3 \int_{-\infty}^{\infty} \left[ \frac{\sin(k_z R) - k_z R \cos(k_z R)}{(k_z R)^3} \right] k_z \chi''(ck_z) \vec{\mathbb{E}}^{(\mathrm{i})}(k_z) \cdot \vec{\mathbb{E}}^{(\mathrm{i})*}(k_z) \mathrm{d}k_z. \quad \text{(C19)}$$

The presence of the imaginary part $\chi''(\cdot)$ of the susceptibility in the above equation indicates that the cross-term energy (which subtracts from the energy of the wave-packet) accounts for both the energy that is permanently absorbed by the particle, as well as the energy that is temporarily absorbed but eventually radiated back into the surrounding medium.

## Appendix D

### Linear momentum of the electromagnetic field radiated by the spherical dipole

The linear momentum of the radiated field at $t_0 > T$ is given by

$$\mathscr{p}(t_0) = \int_{-\infty}^{\infty} c^{-2}\boldsymbol{E}(\boldsymbol{r},t_0)\times\boldsymbol{H}(\boldsymbol{r},t_0)\mathrm{d}\boldsymbol{r}$$

$$= \frac{\mu_0}{2\pi^3}\int_{-\infty}^{\infty}\left[\frac{\sin(kR)-kR\cos(kR)}{k^3}\right]^2\int_{-\infty}^{\infty}\frac{(\omega/c)^2\vec{\mathbb{P}}(\omega)-[\vec{\mathbb{P}}(\omega)\cdot\boldsymbol{k}]\boldsymbol{k}}{k^2-(\omega/c)^2}\exp(-\mathrm{i}\omega t_0)\,\mathrm{d}\omega$$

$$\times\int_{-\infty}^{\infty}\frac{-\boldsymbol{k}\times\omega\vec{\mathbb{P}}(\omega)}{k^2-(\omega/c)^2}\exp(-\mathrm{i}\omega t_0)\mathrm{d}\omega\,\mathrm{d}\boldsymbol{k}$$

$$= -\frac{2\mathrm{i}c}{\pi\varepsilon_0}\int_{-\infty}^{\infty}\left(\frac{1}{k}\right)\left[\frac{\sin(kR)-kR\cos(kR)}{k^3}\right]^2$$

$$\times\left[\sin(ckt_0)\left\{k^2\vec{\mathbb{P}}_R(ck)-[\vec{\mathbb{P}}_R(ck)\cdot\boldsymbol{k}]\boldsymbol{k}\right\}-\cos(ckt_0)\left\{k^2\vec{\mathbb{P}}_I(ck)-[\vec{\mathbb{P}}_I(ck)\cdot\boldsymbol{k}]\boldsymbol{k}\right\}\right]$$

$$\times\left\{\boldsymbol{k}\times[\cos(ckt_0)\vec{\mathbb{P}}_R(ck)+\sin(ckt_0)\vec{\mathbb{P}}_I(ck)]\right\}\mathrm{d}\boldsymbol{k}. \quad \text{(D1)}$$

The integrand of the above equation may be further simplified, as follows:

$$\left\{\sin(ckt_0)[k^2\vec{\mathbb{P}}_R-(\vec{\mathbb{P}}_R\cdot\boldsymbol{k})\boldsymbol{k}]-\cos(ckt_0)[k^2\vec{\mathbb{P}}_I-(\vec{\mathbb{P}}_I\cdot\boldsymbol{k})\boldsymbol{k}]\right\}\times\left\{\boldsymbol{k}\times[\cos(ckt_0)\vec{\mathbb{P}}_R+\sin(ckt_0)\vec{\mathbb{P}}_I]\right\}$$

$$= [k^2\vec{\mathbb{P}}_R-(\vec{\mathbb{P}}_R\cdot\boldsymbol{k})\boldsymbol{k}]\times\left\{\boldsymbol{k}\times[½\sin(2ckt_0)\vec{\mathbb{P}}_R+\sin^2(ckt_0)\vec{\mathbb{P}}_I]\right\}$$

$$-[k^2\vec{\mathbb{P}}_I-(\vec{\mathbb{P}}_I\cdot\boldsymbol{k})\boldsymbol{k}]\times\left\{\boldsymbol{k}\times[\cos^2(ckt_0)\vec{\mathbb{P}}_R+½\sin(2ckt_0)\vec{\mathbb{P}}_I]\right\}$$

$$= [k^2\vec{\mathbb{P}}_R-(\vec{\mathbb{P}}_R\cdot\boldsymbol{k})\boldsymbol{k}]\cdot\left[½\sin(2ckt_0)\vec{\mathbb{P}}_R+\sin^2(ckt_0)\vec{\mathbb{P}}_I\right]\boldsymbol{k}$$

$$-\left\{[k^2\vec{\mathbb{P}}_R-(\vec{\mathbb{P}}_R\cdot\boldsymbol{k})\boldsymbol{k}]\cdot\boldsymbol{k}\right\}\left[½\sin(2ckt_0)\vec{\mathbb{P}}_R+\sin^2(ckt_0)\vec{\mathbb{P}}_I\right]$$

$$-[k^2\vec{\mathbb{P}}_I-(\vec{\mathbb{P}}_I\cdot\boldsymbol{k})\boldsymbol{k}]\cdot\left[\cos^2(ckt_0)\vec{\mathbb{P}}_R+½\sin(2ckt_0)\vec{\mathbb{P}}_I\right]\boldsymbol{k}$$

$$+\left\{[k^2\vec{\mathbb{P}}_I-(\vec{\mathbb{P}}_I\cdot\boldsymbol{k})\boldsymbol{k}]\cdot\boldsymbol{k}\right\}\left[\cos^2(ckt_0)\vec{\mathbb{P}}_R+½\sin(2ckt_0)\vec{\mathbb{P}}_I\right]$$

$$= ½\sin(2ckt_0)\left[k^2(\vec{\mathbb{P}}_R\cdot\vec{\mathbb{P}}_R)-(\boldsymbol{k}\cdot\vec{\mathbb{P}}_R)^2\right]\boldsymbol{k}+\sin^2(ckt_0)\left[k^2(\vec{\mathbb{P}}_R\cdot\vec{\mathbb{P}}_I)-(\boldsymbol{k}\cdot\vec{\mathbb{P}}_R)(\boldsymbol{k}\cdot\vec{\mathbb{P}}_I)\right]\boldsymbol{k}$$

$$-\cos^2(ckt_0)\left[k^2(\vec{\mathbb{P}}_R\cdot\vec{\mathbb{P}}_I)-(\boldsymbol{k}\cdot\vec{\mathbb{P}}_R)(\boldsymbol{k}\cdot\vec{\mathbb{P}}_I)\right]\boldsymbol{k}-½\sin(2ckt_0)\left[k^2(\vec{\mathbb{P}}_I\cdot\vec{\mathbb{P}}_I)-(\boldsymbol{k}\cdot\vec{\mathbb{P}}_I)^2\right]\boldsymbol{k}$$

$$= \left\{½\sin(2ckt_0)\left[(\boldsymbol{k}\times\vec{\mathbb{P}}_R)\cdot(\boldsymbol{k}\times\vec{\mathbb{P}}_R)-(\boldsymbol{k}\times\vec{\mathbb{P}}_I)\cdot(\boldsymbol{k}\times\vec{\mathbb{P}}_I)\right]-\cos(2ckt_0)\left[(\boldsymbol{k}\times\vec{\mathbb{P}}_R)\cdot(\boldsymbol{k}\times\vec{\mathbb{P}}_I)\right]\right\}\boldsymbol{k}. \quad \text{(D2)}$$

Equation (D1) may thus be written as

$$\mathscr{p}(t_0) = -\frac{2\mathrm{i}c}{\pi\varepsilon_0}\int_{-\infty}^{\infty}\left(\frac{1}{k}\right)\left[\frac{\sin(kR)-kR\cos(kR)}{k^3}\right]^2\left\{½\sin(2ckt_0)\,[(\boldsymbol{k}\times\vec{\mathbb{P}}_R)\cdot(\boldsymbol{k}\times\vec{\mathbb{P}}_R)-(\boldsymbol{k}\times\vec{\mathbb{P}}_I)\cdot(\boldsymbol{k}\times\vec{\mathbb{P}}_I)]\right.$$

Antisymmetric integrand with respect to a sign-change of $\boldsymbol{k}$. →

$$\left.-\cos(2ckt_0)\,[(\boldsymbol{k}\times\vec{\mathbb{P}}_R)\cdot(\boldsymbol{k}\times\vec{\mathbb{P}}_I)]\right\}\boldsymbol{k}\mathrm{d}\boldsymbol{k} = 0. \quad \text{(D3)}$$

The linear momentum of the radiated field (integrated over the entire $xyz$ space) is thus seen to be zero for $t_0 > T$. This is consistent with the fact that the EM force exerted on the dipole by the radiated field, when integrated over all time, is found to be zero, as shown below.

$$\boldsymbol{f}_{\text{total}} = \int_{-\infty}^{\infty}\int_{t=-\infty}^{\infty}[(\boldsymbol{P}\cdot\boldsymbol{\nabla})\boldsymbol{E}+\partial_t\boldsymbol{P}\times\mu_0\boldsymbol{H}]\mathrm{d}\boldsymbol{r}\mathrm{d}t$$

··· continued on the next page

$$= \frac{1}{(4\pi^3)^2\varepsilon_0}\int_{-\infty}^{\infty}\left[\frac{\sin(kR)-kR\cos(kR)}{k^3}\right][\vec{\mathbb{P}}(\omega)\cdot \mathrm{i}\boldsymbol{k}']\exp[\mathrm{i}(\boldsymbol{k}\cdot\boldsymbol{r}-\omega t)]$$

$$\times \frac{\{(\omega'/c)^2\vec{\mathbb{P}}(\omega') - [\vec{\mathbb{P}}(\omega')\cdot\boldsymbol{k}']\boldsymbol{k}'\}[\sin(k'R)-k'R\cos(k'R)]}{k'^3[k'^2-(\omega'/c)^2]}\exp[\mathrm{i}(\boldsymbol{k}'\cdot\boldsymbol{r}-\omega' t)]\,\mathrm{d}\boldsymbol{k}\mathrm{d}\omega\mathrm{d}\boldsymbol{k}'\mathrm{d}\omega'\mathrm{d}\boldsymbol{r}\mathrm{d}t$$

$$+\frac{\mu_0}{(4\pi^3)^2}\int_{-\infty}^{\infty}\frac{-\mathrm{i}\omega\vec{\mathbb{P}}(\omega)[\sin(kR)-kR\cos(kR)]}{k^3}\exp[\mathrm{i}(\boldsymbol{k}\cdot\boldsymbol{r}-\omega t)]$$

$$\times\frac{\boldsymbol{k}'\times\omega'\vec{\mathbb{P}}(\omega')[\sin(k'R)-k'R\cos(k'R)]}{k'^3[k'^2-(\omega'/c)^2]}\exp[\mathrm{i}(\boldsymbol{k}'\cdot\boldsymbol{r}-\omega' t)]\,\mathrm{d}\boldsymbol{k}\mathrm{d}\omega\mathrm{d}\boldsymbol{k}'\mathrm{d}\omega'\mathrm{d}\boldsymbol{r}\mathrm{d}t$$

$$=\frac{-\mathrm{i}(2\pi)^4}{(4\pi^3)^2\varepsilon_0}\int_{-\infty}^{\infty}\left[\frac{\sin(kR)-kR\cos(kR)}{k^3}\right]^2\frac{[\vec{\mathbb{P}}(\omega)\cdot\boldsymbol{k}]\{(\omega/c)^2\vec{\mathbb{P}}^*(\omega)-[\vec{\mathbb{P}}^*(\omega)\cdot\boldsymbol{k}]\boldsymbol{k}\}}{k^2-(\omega/c)^2}\mathrm{d}\boldsymbol{k}\mathrm{d}\omega$$

$$-\frac{\mathrm{i}(2\pi)^4\mu_0}{(4\pi^3)^2}\int_{-\infty}^{\infty}\left[\frac{\sin(kR)-kR\cos(kR)}{k^3}\right]^2\frac{\omega\vec{\mathbb{P}}(\omega)\times[\boldsymbol{k}\times\omega\vec{\mathbb{P}}^*(\omega)]}{k^2-(\omega/c)^2}\mathrm{d}\boldsymbol{k}\mathrm{d}\omega$$

$$=\frac{-\mathrm{i}}{\pi^2\varepsilon_0}\int_{-\infty}^{\infty}\left[\frac{\sin(kR)-kR\cos(kR)}{k^3}\right]^2$$

$$\times\frac{(\omega/c)^2\{[\vec{\mathbb{P}}(\omega)\cdot\boldsymbol{k}]\vec{\mathbb{P}}^*(\omega)+[\vec{\mathbb{P}}(\omega)\cdot\vec{\mathbb{P}}^*(\omega)]\boldsymbol{k}-[\vec{\mathbb{P}}(\omega)\cdot\boldsymbol{k}]\vec{\mathbb{P}}^*(\omega)\}-[\vec{\mathbb{P}}(\omega)\cdot\boldsymbol{k}][\vec{\mathbb{P}}^*(\omega)\cdot\boldsymbol{k}]\boldsymbol{k}}{k^2-(\omega/c)^2}\mathrm{d}\boldsymbol{k}\mathrm{d}\omega$$

$$=\frac{\mathrm{i}}{\pi^2\varepsilon_0}\int_{-\infty}^{\infty}\left[\frac{\sin(kR)-kR\cos(kR)}{k^3}\right]^2\frac{[\vec{\mathbb{P}}(\omega)\cdot\boldsymbol{k}][\vec{\mathbb{P}}^*(\omega)\cdot\boldsymbol{k}]-(\omega/c)^2\vec{\mathbb{P}}(\omega)\cdot\vec{\mathbb{P}}^*(\omega)}{k^2-(\omega/c)^2}\boldsymbol{k}\mathrm{d}\boldsymbol{k}\mathrm{d}\omega$$

$$=\frac{\mathrm{i}}{\pi^2\varepsilon_0}\int_{-\infty}^{\infty}\left[\frac{\sin(kR)-kR\cos(kR)}{k^3}\right]^2\left\{\vec{\mathbb{P}}(\omega)\cdot\vec{\mathbb{P}}^*(\omega)-\frac{[\boldsymbol{k}\times\vec{\mathbb{P}}(\omega)]\cdot[\boldsymbol{k}\times\vec{\mathbb{P}}^*(\omega)]}{k^2-(\omega/c)^2}\right\}\boldsymbol{k}\mathrm{d}\boldsymbol{k}\mathrm{d}\omega=0. \qquad \text{(D4)}$$

Given the *antisymmetry* of the integrand with respect to a sign-change of $\boldsymbol{k}$, the overall integral vanishes. Note that, in these calculations, we have neglected to include the pair of $\delta$-functions that appear in Eq.(14). However, since the integrand is *symmetric* with respect to a sign-change of $\omega$, the contributions of the two $\delta$-functions cancel out.

**Cross-terms between incident beam and the radiation of the induced dipole $\vec{\mathbb{P}}(\boldsymbol{\omega})=\boldsymbol{\varepsilon_0\chi(\omega)E^{(\mathrm{i})}(\omega)}$:**

$$\boldsymbol{\wp}_{\text{cross}}(t_0)=c^{-2}\int_{-\infty}^{\infty}[\boldsymbol{E}^{(\mathrm{i})}(\boldsymbol{r},t_0)\times\boldsymbol{H}^{(\mathrm{r})}(\boldsymbol{r},t_0)-\boldsymbol{H}^{(\mathrm{i})}(\boldsymbol{r},t_0)\times\boldsymbol{E}^{(\mathrm{r})}(\boldsymbol{r},t_0)]\mathrm{d}\boldsymbol{r}$$

$$=\frac{1}{(2\pi)^5c^2}\int_{-\infty}^{\infty}[\vec{\mathbb{E}}^{(\mathrm{i})}(\omega)\times\vec{\mathbb{H}}^{(\mathrm{r})}(\boldsymbol{k}',\omega')-\vec{\mathbb{H}}^{(\mathrm{i})}(\omega)\times\vec{\mathbb{E}}^{(\mathrm{r})}(\boldsymbol{k}',\omega')]$$

$$\times\exp\{\mathrm{i}[k_x'x+k_y'y+(k_z'+\omega/c)z]\}\exp[-\mathrm{i}(\omega+\omega')t_0]\,\mathrm{d}\omega\mathrm{d}\boldsymbol{k}'\mathrm{d}\omega'\mathrm{d}\boldsymbol{r}$$

$$=\frac{(2\pi)^3}{8\pi^4c^2}\int_{-\infty}^{\infty}\frac{\sin(k'R)-k'R\cos(k'R)}{k'^3[k'^2-(\omega'/c)^2]}\{\vec{\mathbb{E}}^{(\mathrm{i})}(\omega)\times[\boldsymbol{k}'\times\omega'\vec{\mathbb{P}}(\omega')]-[c\hat{\boldsymbol{z}}\times\vec{\mathbb{E}}^{(\mathrm{i})}(\omega)]$$

$$\times\{(\omega'/c)^2\vec{\mathbb{P}}(\omega')-[\vec{\mathbb{P}}(\omega')\cdot\boldsymbol{k}']\boldsymbol{k}'\}\}\delta(k_x')\delta(k_y')\delta(k_z'+\omega/c)$$

$$\times\exp[-\mathrm{i}(\omega+\omega')t_0]\,\mathrm{d}\omega\mathrm{d}\boldsymbol{k}'\mathrm{d}\omega'$$

$$=\frac{1}{\pi c^2}\int_{-\infty}^{\infty}\frac{\sin(R\omega/c)-(R\omega/c)\cos(R\omega/c)}{(\omega/c)^3[(\omega/c)^2-(\omega'/c)^2]}\{-(\omega\omega'/c)\vec{\mathbb{E}}^{(\mathrm{i})}(\omega)\times[\hat{\boldsymbol{z}}\times\vec{\mathbb{P}}(\omega')] \quad \leftarrow \boxed{\boldsymbol{E}^{(\mathrm{i})}(\omega)\cdot\hat{\boldsymbol{z}}=0}$$

$$-(\omega'^2/c)[\hat{\boldsymbol{z}}\times\vec{\mathbb{E}}^{(\mathrm{i})}(\omega)]\times\{\vec{\mathbb{P}}(\omega')-\cancelto{0}{[\vec{\mathbb{P}}(\omega')\cdot\hat{\boldsymbol{z}}]\hat{\boldsymbol{z}}}\}\}\times\exp[-\mathrm{i}(\omega+\omega')t_0]\,\mathrm{d}\omega\mathrm{d}\omega'$$

$$= \frac{R^3 \hat{\boldsymbol{z}}}{\pi c^2} \int_{-\infty}^{\infty} \left[\frac{\sin(R\omega/c) - (R\omega/c)\cos(R\omega/c)}{(R\omega/c)^3}\right] \left[\frac{(\omega'/c)(\omega' - \omega)\vec{\mathbb{E}}^{(\mathrm{i})}(\omega) \cdot \vec{\mathbb{P}}(\omega')}{(\omega/c)^2 - (\omega'/c)^2}\right]$$

$$\times \exp[-\mathrm{i}(\omega + \omega')t_0]\,\mathrm{d}\omega\mathrm{d}\omega'. \tag{D5}$$

Considering that

$$\int_{-\infty}^{\infty} \frac{(\omega'/c)(\omega' - \omega)\vec{\mathbb{E}}^{(\mathrm{i})}(\omega) \cdot \vec{\mathbb{P}}(\omega')}{(\omega/c)^2 - (\omega'/c)^2} \exp(-\mathrm{i}\omega' t_0)\,\mathrm{d}\omega'$$

$$= 2\pi(c/\omega)\vec{\mathbb{E}}^{(\mathrm{i})}(\omega) \cdot \{\omega^2[\sin(\omega t_0)\,\vec{\mathbb{P}}_R(\omega) - \cos(\omega t_0)\,\vec{\mathbb{P}}_I(\omega)]$$

$$-\mathrm{i}\omega^2[\cos(\omega t_0)\,\vec{\mathbb{P}}_R(\omega) + \sin(\omega t_0)\,\vec{\mathbb{P}}_I(\omega)]\}$$

$$= -\mathrm{i}2\pi c\omega\vec{\mathbb{E}}^{(\mathrm{i})}(\omega) \cdot [\vec{\mathbb{P}}_R(\omega) - \mathrm{i}\vec{\mathbb{P}}_I(\omega)]\exp(\mathrm{i}\omega t_0)$$

$$= -\mathrm{i}2\pi c\omega\vec{\mathbb{E}}^{(\mathrm{i})}(\omega) \cdot \varepsilon_0\chi^*(\omega)\vec{\mathbb{E}}^{(\mathrm{i})*}(\omega)\exp(\mathrm{i}\omega t_0). \tag{D6}$$

We will have

$$\boldsymbol{\wp}_{\text{cross}}(t_0) = -\frac{\mathrm{i}2R^3\hat{\boldsymbol{z}}}{c^2 Z_0} \int_{-\infty}^{\infty} \left[\frac{\sin(R\omega/c) - (R\omega/c)\cos(R\omega/c)}{(R\omega/c)^3}\right] \omega\chi^*(\omega)\vec{\mathbb{E}}^{(\mathrm{i})}(\omega) \cdot \vec{\mathbb{E}}^{(\mathrm{i})*}(\omega)\mathrm{d}\omega$$

$$= -\frac{2R^3\hat{\boldsymbol{z}}}{c^2 Z_0} \int_{-\infty}^{\infty} \left[\frac{\sin(R\omega/c) - (R\omega/c)\cos(R\omega/c)}{(R\omega/c)^3}\right] \mathrm{Im}[\omega\chi(\omega)]\vec{\mathbb{E}}^{(\mathrm{i})}(\omega) \cdot \vec{\mathbb{E}}^{(\mathrm{i})*}(\omega)\mathrm{d}\omega$$

$$\boxed{R\omega/c \ll 1} \rightarrow \cong -\frac{(4\pi/3)R^3\hat{\boldsymbol{z}}}{\pi c^2 Z_0} \int_0^{\infty} \mathrm{Im}[\omega\chi(\omega)]\vec{\mathbb{E}}^{(\mathrm{i})}(\omega) \cdot \vec{\mathbb{E}}^{(\mathrm{i})*}(\omega)\mathrm{d}\omega. \tag{D7}$$

The above momentum is deducted from the momentum of the incident wave-packet, which has continued to propagate beyond the particle. The momentum that is thus lost to the electromagnetic wave is picked up (as mechanical momentum) by the particle, partly due to absorption (if the damping coefficient $\gamma$ happens to be non-zero), and in part due to scattering.

## Appendix E

### Angular momentum of the electromagnetic field radiated by the spherical dipole

The angular momentum of the radiated field at $t_0 > T$ is given by

$$\boldsymbol{\mathcal{L}}(t_0) = c^{-2}\int_{-\infty}^{\infty} \boldsymbol{r} \times [\boldsymbol{E}(\boldsymbol{r}, t_0) \times \boldsymbol{H}(\boldsymbol{r}, t_0)]\mathrm{d}\boldsymbol{r}$$

$$= \frac{\mu_0}{16\pi^6}\int_{-\infty}^{\infty} \left\{\int_{-\infty}^{\infty} \boldsymbol{r}\exp[\mathrm{i}(\boldsymbol{k}+\boldsymbol{k}')\cdot\boldsymbol{r}]\,\mathrm{d}\boldsymbol{r}\right\} \times \left\{\frac{\{(\omega/c)^2\vec{\mathbb{P}}(\omega) - [\vec{\mathbb{P}}(\omega)\cdot\boldsymbol{k}]\boldsymbol{k}\}[\sin(kR) - kR\cos(kR)]}{k^3[k^2-(\omega/c)^2]}\right.$$

$$\left. \times \frac{\boldsymbol{k}' \times \omega'\vec{\mathbb{P}}(\omega')[\sin(k'R) - k'R\cos(k'R)]}{k'^3[k'^2-(\omega'/c)^2]}\right\}\exp[-\mathrm{i}(\omega+\omega')t_0]\,\mathrm{d}\boldsymbol{k}\mathrm{d}\boldsymbol{k}'\mathrm{d}\omega\mathrm{d}\omega'. \quad \text{(E1)}$$

Invoking the identities $\int_{-\infty}^{\infty}\exp(\mathrm{i}kx)\,\mathrm{d}x = 2\pi\delta(k)$ and $\int_{-\infty}^{\infty} x\exp(\mathrm{i}kx)\,\mathrm{d}x = -\mathrm{i}2\pi\delta'(k)$, the inner integral in the above equation may be evaluated as follows:

$$\int_{-\infty}^{\infty} \boldsymbol{r}\exp[\mathrm{i}(\boldsymbol{k}+\boldsymbol{k}')\cdot\boldsymbol{r}]\,\mathrm{d}\boldsymbol{r} = -\mathrm{i}(2\pi)^3\big[\delta'(k_x+k_x')\delta(k_y+k_y')\delta(k_z+k_z')\hat{\boldsymbol{x}}$$

$$+\delta(k_x+k_x')\delta'(k_y+k_y')\delta(k_z+k_z')\hat{\boldsymbol{y}} + \delta(k_x+k_x')\delta(k_y+k_y')\delta'(k_z+k_z')\hat{\boldsymbol{z}}\big]. \quad \text{(E2)}$$

The notation will be enormously simplified if we now define the following functions:

$$\boldsymbol{\mathcal{D}}(\boldsymbol{k}+\boldsymbol{k}') = \mathrm{i}(2\pi)^{-3}\int_{-\infty}^{\infty}\boldsymbol{r}\exp[\mathrm{i}(\boldsymbol{k}+\boldsymbol{k}')\cdot\boldsymbol{r}]\,\mathrm{d}\boldsymbol{r}, \quad \text{(E3)}$$

$$\boldsymbol{\mathcal{R}}(\boldsymbol{k},\omega) = \frac{\{(\omega/c)^2\vec{\mathbb{P}}(\omega) - [\vec{\mathbb{P}}(\omega)\cdot\boldsymbol{k}]\boldsymbol{k}\}[\sin(kR)-kR\cos(kR)]}{k^3[k^2-(\omega/c)^2]}, \quad \text{(E4)}$$

$$\boldsymbol{\mathcal{S}}(\boldsymbol{k}',\omega') = \frac{\boldsymbol{k}'\times\omega'\vec{\mathbb{P}}(\omega')[\sin(k'R)-k'R\cos(k'R)]}{k'^3[k'^2-(\omega'/c)^2]}. \quad \text{(E5)}$$

Substituting the above functions into Eq.(E1), and using the sifting property of the various $\delta$-functions and their derivatives appearing in Eq.(E2), namely, $\int_{-\infty}^{\infty}\delta(k+k')f(k)\mathrm{d}k = f(-k')$ and $\int_{-\infty}^{\infty}\delta'(k+k')f(k)\mathrm{d}k = -f'(-k')$, we find

$$\boldsymbol{\mathcal{L}}(t_0) = \frac{-\mathrm{i}\mu_0}{2\pi^3}\int_{-\infty}^{\infty}\boldsymbol{\mathcal{D}}(\boldsymbol{k}+\boldsymbol{k}')\times[\boldsymbol{\mathcal{R}}(\boldsymbol{k},\omega)\times\boldsymbol{\mathcal{S}}(\boldsymbol{k}',\omega')]\exp[-\mathrm{i}(\omega+\omega')t_0]\,\mathrm{d}\boldsymbol{k}\mathrm{d}\boldsymbol{k}'\mathrm{d}\omega\mathrm{d}\omega'$$

$$= \frac{-\mathrm{i}\mu_0}{2\pi^3}\int_{-\infty}^{\infty}[(\boldsymbol{\mathcal{D}}\cdot\boldsymbol{\mathcal{S}})\boldsymbol{\mathcal{R}} - (\boldsymbol{\mathcal{D}}\cdot\boldsymbol{\mathcal{R}})\boldsymbol{\mathcal{S}}]\exp[-\mathrm{i}(\omega+\omega')t_0]\,\mathrm{d}\boldsymbol{k}\mathrm{d}\boldsymbol{k}'\mathrm{d}\omega\mathrm{d}\omega'$$

$$= \frac{\mathrm{i}\mu_0}{2\pi^3}\Big\{\int_{-\infty}^{\infty}(\boldsymbol{\nabla}_{k'}\cdot\boldsymbol{\mathcal{S}})|_{\boldsymbol{k}'=-\boldsymbol{k}}\boldsymbol{\mathcal{R}}(\boldsymbol{k},\omega)\exp[-\mathrm{i}(\omega+\omega')t_0]\,\mathrm{d}\boldsymbol{k}\mathrm{d}\omega\mathrm{d}\omega'$$

$$-\int_{-\infty}^{\infty}(\boldsymbol{\nabla}_k\cdot\boldsymbol{\mathcal{R}})|_{\boldsymbol{k}=-\boldsymbol{k}'}\boldsymbol{\mathcal{S}}(\boldsymbol{k}',\omega')\exp[-\mathrm{i}(\omega+\omega')t_0]\,\mathrm{d}\boldsymbol{k}'\mathrm{d}\omega\mathrm{d}\omega'\Big\}. \quad \text{(E6)}$$

It is easy to show that $\boldsymbol{\nabla}_{k'}\cdot\boldsymbol{\mathcal{S}}(\boldsymbol{k}',\omega')$ appearing in the above equation equals zero, that is,

$$\boldsymbol{\nabla}_{k'}\cdot\boldsymbol{\mathcal{S}}(\boldsymbol{k}',\omega') = \boldsymbol{\nabla}_{k'}\cdot\left\{\frac{\boldsymbol{k}'\times\omega'\vec{\mathbb{P}}(\omega')[\sin(k'R)-k'R\cos(k'R)]}{k'^3[k'^2-(\omega'/c)^2]}\right\}$$

$$= \big[\boldsymbol{k}'\times\omega'\vec{\mathbb{P}}(\omega')\big]\cdot\boldsymbol{\nabla}_{k'}\left\{\frac{\sin(k'R)-k'R\cos(k'R)}{k'^3[k'^2-(\omega'/c)^2]}\right\} \quad \leftarrow \boxed{\boldsymbol{\nabla}_{k'}\cdot[\boldsymbol{k}'\times\omega'\vec{\mathbb{P}}(\omega')] = 0.}$$

$$= \big[\boldsymbol{k}'\times\omega'\vec{\mathbb{P}}(\omega')\big]\cdot\hat{\boldsymbol{k}}'\,\partial_{k'}\left\{\frac{\sin(k'R)-k'R\cos(k'R)}{k'^3[k'^2-(\omega'/c)^2]}\right\} = 0. \quad \text{(E7)}$$

As for the remaining term in Eq.(E6), we write

$$\boldsymbol{\nabla}_k\cdot\boldsymbol{\mathcal{R}}(\boldsymbol{k},\omega) = \boldsymbol{\nabla}_k\cdot\frac{\{(\omega/c)^2\vec{\mathbb{P}}(\omega) - [\vec{\mathbb{P}}(\omega)\cdot\boldsymbol{k}]\boldsymbol{k}\}[\sin(kR)-kR\cos(kR)]}{k^3[k^2-(\omega/c)^2]}$$

…continued on the next page

$$= \{(\omega/c)^2\vec{\mathbb{P}}(\omega) - [\vec{\mathbb{P}}(\omega)\cdot\boldsymbol{k}]\boldsymbol{k}\}\cdot\boldsymbol{\nabla}_k\left\{\frac{\sin(kR)-kR\cos(kR)}{k^3[k^2-(\omega/c)^2]}\right\}$$

$$+\left\{\frac{\sin(kR)-kR\cos(kR)}{k^3[k^2-(\omega/c)^2]}\right\}\boldsymbol{\nabla}_k\cdot\{(\omega/c)^2\vec{\mathbb{P}}(\omega) - [\vec{\mathbb{P}}(\omega)\cdot\boldsymbol{k}]\boldsymbol{k}\}$$

$$= \{(\omega/c)^2\vec{\mathbb{P}}(\omega) - [\vec{\mathbb{P}}(\omega)\cdot\boldsymbol{k}]\boldsymbol{k}\}\cdot\hat{\boldsymbol{k}}\,\partial_k\left\{\frac{\sin(kR)-kR\cos(kR)}{k^3[k^2-(\omega/c)^2]}\right\}$$

$$-\left\{\frac{\sin(kR)-kR\cos(kR)}{k^3[k^2-(\omega/c)^2]}\right\}\boldsymbol{\nabla}_k\cdot\{[\vec{\mathbb{P}}(\omega)\cdot\boldsymbol{k}]\boldsymbol{k}\}$$

$$= -\left(k^{-1}[k^2-(\omega/c)^2]\partial_k\left\{\frac{\sin(kR)-kR\cos(kR)}{k^3[k^2-(\omega/c)^2]}\right\} + \frac{4[\sin(kR)-kR\cos(kR)]}{k^3[k^2-(\omega/c)^2]}\right)\vec{\mathbb{P}}(\omega)\cdot\boldsymbol{k}. \qquad \text{(E8)}$$

The derivative with respect to $k$ appearing in Eq.(E8) is straightforwardly computed, as follows:

$$\partial_k\left\{\frac{\sin(kR)-kR\cos(kR)}{k^3[k^2-(\omega/c)^2]}\right\} = \frac{R^2\sin(kR)}{k^2[k^2-(\omega/c)^2]} - \frac{3[\sin(kR)-kR\cos(kR)]}{k^4[k^2-(\omega/c)^2]} - \frac{2[\sin(kR)-kR\cos(kR)]}{k^2[k^2-(\omega/c)^2]^2}. \qquad \text{(E9)}$$

Substitution into Eq.(E8) now yields

$$\boldsymbol{\nabla}_k\cdot\boldsymbol{\mathcal{R}}(\boldsymbol{k},\omega) = -\left\{\frac{R^2\sin(kR)}{k^3} - \frac{3[\sin(kR)-kR\cos(kR)]}{k^5} + \frac{2[\sin(kR)-kR\cos(kR)]}{k^3[k^2-(\omega/c)^2]}\right\}\vec{\mathbb{P}}(\omega)\cdot\boldsymbol{k}. \qquad \text{(E10)}$$

Returning to Eq.(E6) with the aid of Eqs.(E7) and (E10), we now find

$$\boldsymbol{\mathcal{L}}(t_0) = \frac{-\mathrm{i}\mu_0}{2\pi^3}\int_{-\infty}^{\infty}\left\{\left(\frac{R^2\sin(k'R)}{k'^3} - \frac{3[\sin(k'R)-k'R\cos(k'R)]}{k'^5}\right)\boldsymbol{k}'\cdot\underbrace{\int_{-\infty}^{\infty}\vec{\mathbb{P}}(\omega)\exp(-\mathrm{i}\omega t_0)\,\mathrm{d}\omega}_{2\pi\boldsymbol{P}(t_0)\,=\,0.}\right.$$

$$\left.+2\left[\frac{\sin(k'R)-k'R\cos(k'R)}{k'^3}\right]\boldsymbol{k}'\cdot\int_{-\infty}^{\infty}\frac{\vec{\mathbb{P}}(\omega)\exp(-\mathrm{i}\omega t_0)}{k'^2-(\omega/c)^2}\mathrm{d}\omega\right\}$$

$$\left\{\left[\frac{\sin(k'R)-k'R\cos(k'R)}{k'^3}\right]\boldsymbol{k}'\times\int_{-\infty}^{\infty}\frac{\omega'\vec{\mathbb{P}}(\omega')\exp(-\mathrm{i}\omega' t_0)}{k'^2-(\omega'/c)^2}\mathrm{d}\omega'\right\}\mathrm{d}\boldsymbol{k}'$$

$$= \frac{-\mathrm{i}\mu_0}{\pi^3}\int_{-\infty}^{\infty}\left[\frac{\sin(k'R)-k'R\cos(k'R)}{k'^3}\right]^2\{2\pi c\hat{\boldsymbol{k}}'\cdot[\sin(ck't_0)\vec{\mathbb{P}}_R(ck') - \cos(ck't_0)\vec{\mathbb{P}}_I(ck')]\}$$

$$\{\mathrm{i}2\pi c\hat{\boldsymbol{k}}'\times[\cos(ck't_0)\,ck'\vec{\mathbb{P}}_R(ck') + \sin(ck't_0)\,ck'\vec{\mathbb{P}}_I(ck')]\}\mathrm{d}\boldsymbol{k}'$$

$$= \frac{4R^6c}{\pi\varepsilon_0}\int_{-\infty}^{\infty}\left[\frac{\sin(kR)-kR\cos(kR)}{(kR)^3}\right]^2\left(\frac{1}{k}\right)\{\boldsymbol{k}\cdot[\sin(ckt_0)\vec{\mathbb{P}}_R(ck) - \cos(ckt_0)\vec{\mathbb{P}}_I(ck)]\}$$

$$\{\boldsymbol{k}\times[\cos(ckt_0)\vec{\mathbb{P}}_R(ck) + \sin(ckt_0)\vec{\mathbb{P}}_I(ck)]\}\mathrm{d}\boldsymbol{k}. \qquad \text{(E11)}$$

← substitute $\boldsymbol{k}$ for $\boldsymbol{k}'$

Let us now assume that $\vec{\mathbb{P}}_R(\omega) = \mathbb{P}_R(\omega)\hat{\boldsymbol{x}}$ and $\vec{\mathbb{P}}_I(\omega) = \mathbb{P}_I(\omega)\hat{\boldsymbol{y}}$. The total radiated angular momentum at $t = t_0$ becomes

$$\boldsymbol{\mathcal{L}}(t_0) = \frac{4R^6c}{\pi\varepsilon_0}\int_{-\infty}^{\infty}\left(\frac{1}{k}\right)\left[\frac{\sin(kR)-kR\cos(kR)}{(kR)^3}\right]^2[k_x\mathbb{P}_R(ck)\sin(ckt_0) - k_y\mathbb{P}_I(ck)\cos(ckt_0)]$$

$$[(k_z\hat{\boldsymbol{y}} - k_y\hat{\boldsymbol{z}})\mathbb{P}_R(ck)\cos(ckt_0) + (k_x\hat{\boldsymbol{z}} - k_z\hat{\boldsymbol{x}})\mathbb{P}_I(ck)\sin(ckt_0)]\mathrm{d}\boldsymbol{k}$$

$$= \frac{4R^6c\hat{\boldsymbol{z}}}{\pi\varepsilon_0}\int_{-\infty}^{\infty}\left(\frac{1}{k}\right)\left[\frac{\sin(kR)-kR\cos(kR)}{(kR)^3}\right]^2[k_x^2\sin^2(ckt_0) + k_y^2\cos^2(ckt_0)]\mathbb{P}_R(ck)\mathbb{P}_I(ck)\mathrm{d}\boldsymbol{k}$$

$$= \frac{2R^6c\hat{\boldsymbol{z}}}{\pi\varepsilon_0}\int_{-\infty}^{\infty}\left(\frac{1}{k}\right)\left[\frac{\sin(kR)-kR\cos(kR)}{(kR)^3}\right]^2(k_x^2 + k_y^2)\mathbb{P}_R(ck)\mathbb{P}_I(ck)\mathrm{d}\boldsymbol{k}$$

…continued on the next page

$$= \frac{4R^6 c\hat{\boldsymbol{z}}}{\varepsilon_0} \int_{k=0}^{\infty} \int_{\theta=0}^{\pi} \left[\frac{\sin(kR) - kR\cos(kR)}{(kR)^3}\right]^2 k^3 \sin^3\theta \, \mathbb{P}_R(ck) \mathbb{P}_I(ck) \mathrm{d}\theta \mathrm{d}k$$

← $\int_0^\pi \sin^3\theta \, d\theta = 4/3$

$$= \frac{16\mu_0 R^6 \hat{\boldsymbol{z}}}{3c} \int_0^{\infty} \left[\frac{\sin(R\omega/c) - (R\omega/c)\cos(R\omega/c)}{(R\omega/c)^3}\right]^2 \omega^3 \mathbb{P}_R(\omega) \mathbb{P}_I(\omega) \mathrm{d}\omega. \qquad \text{(E12)}$$

Recall that the energy content of the radiated field under similar circumstances is given by Eq.(C9).

**Angular momentum cross-terms for the one-dimensional wave-packet**: Implicit in the calculations below is the fact that $E_z(\boldsymbol{r}, t) = 0$ and that, therefore, $\mathbb{E}_z(k_z) = 0$ and $\mathbb{P}_z(\omega) = 0$.

$$\boldsymbol{\mathcal{L}}_{\text{cross}}(t_0) = \frac{1}{(2\pi)(4\pi^3)c^2} \int_{-\infty}^{\infty} \boldsymbol{r} \times \left[\vec{\mathbb{E}}^{(\mathrm{i})}(k_z') \times \vec{\mathbb{H}}^{(\mathrm{r})}(\boldsymbol{k}, \omega) + \vec{\mathbb{E}}^{(\mathrm{r})}(\boldsymbol{k}, \omega) \times \vec{\mathbb{H}}^{(\mathrm{i})}(k_z')\right]$$
$$\times \exp[\mathrm{i}k_z'(z - ct_0)] \exp[\mathrm{i}(\boldsymbol{k} \cdot \boldsymbol{r} - \omega t_0)] \, \mathrm{d}\boldsymbol{r} \mathrm{d}k_z' \mathrm{d}\boldsymbol{k} \mathrm{d}\omega. \qquad \text{(E13)}$$

$$\vec{\mathbb{E}}^{(\mathrm{i})}(k_z') \times \vec{\mathbb{H}}^{(\mathrm{r})}(\boldsymbol{k}, \omega) = \vec{\mathbb{E}}^{(\mathrm{i})}(k_z') \times \frac{\boldsymbol{k} \times \omega \vec{\mathbb{P}}(\omega)[\sin(kR) - kR\cos(kR)]}{k^3[k^2 - (\omega/c)^2]}$$
$$= \frac{\sin(kR) - kR\cos(kR)}{k^3[k^2 - (\omega/c)^2]} \omega\left\{[\vec{\mathbb{P}}(\omega) \cdot \vec{\mathbb{E}}^{(\mathrm{i})}(k_z')]\boldsymbol{k} - [\boldsymbol{k} \cdot \vec{\mathbb{E}}^{(\mathrm{i})}(k_z')] \vec{\mathbb{P}}(\omega)\right\}. \qquad \text{(E14)}$$

$$\vec{\mathbb{E}}^{(\mathrm{r})}(\boldsymbol{k}, \omega) \times \vec{\mathbb{H}}^{(\mathrm{i})}(k_z') = \frac{\sin(kR) - kR\cos(kR)}{\varepsilon_0 Z_0 k^3[k^2 - (\omega/c)^2]} \left\{(\omega/c)^2 \vec{\mathbb{P}}(\omega) - [\vec{\mathbb{P}}(\omega) \cdot \boldsymbol{k}]\boldsymbol{k}\right\} \times [\hat{\boldsymbol{z}} \times \vec{\mathbb{E}}^{(\mathrm{i})}(k_z')]$$
$$= \frac{\sin(kR) - kR\cos(kR)}{k^3[k^2 - (\omega/c)^2]} c\left\{(\omega/c)^2 [\vec{\mathbb{P}}(\omega) \cdot \vec{\mathbb{E}}^{(\mathrm{i})}(k_z')]\hat{\boldsymbol{z}} - (\omega/c)^2 \mathbb{P}_z(\omega) \vec{\mathbb{E}}^{(\mathrm{i})}(k_z')\right.$$
$$\left. - [\vec{\mathbb{P}}(\omega) \cdot \boldsymbol{k}][\boldsymbol{k} \cdot \vec{\mathbb{E}}^{(\mathrm{i})}(k_z')]\hat{\boldsymbol{z}} + k_z [\vec{\mathbb{P}}(\omega) \cdot \boldsymbol{k}] \vec{\mathbb{E}}^{(\mathrm{i})}(k_z')\right\}. \qquad \text{(E15)}$$

$$\boldsymbol{r} \times \left[\vec{\mathbb{E}}^{(\mathrm{i})}(k_z') \times \vec{\mathbb{H}}^{(\mathrm{r})}(\boldsymbol{k}, \omega) + \vec{\mathbb{E}}^{(\mathrm{r})}(\boldsymbol{k}, \omega) \times \vec{\mathbb{H}}^{(\mathrm{i})}(k_z')\right] = \frac{\sin(kR) - kR\cos(kR)}{k^3[k^2 - (\omega/c)^2]} \left\{\omega[\vec{\mathbb{P}}(\omega) \cdot \vec{\mathbb{E}}^{(\mathrm{i})}(k_z')](\boldsymbol{r} \times \boldsymbol{k})\right.$$
$$- \omega[\boldsymbol{k} \cdot \vec{\mathbb{E}}^{(\mathrm{i})}(k_z')][\boldsymbol{r} \times \vec{\mathbb{P}}(\omega)] + c\left\{(\omega/c)^2 [\vec{\mathbb{P}}(\omega) \cdot \vec{\mathbb{E}}^{(\mathrm{i})}(k_z')] - [\boldsymbol{k} \cdot \vec{\mathbb{P}}(\omega)][\boldsymbol{k} \cdot \vec{\mathbb{E}}^{(\mathrm{i})}(k_z')]\right\}(\boldsymbol{r} \times \hat{\boldsymbol{z}})$$
$$\left. + ck_z [\boldsymbol{k} \cdot \vec{\mathbb{P}}(\omega)][\boldsymbol{r} \times \vec{\mathbb{E}}^{(\mathrm{i})}(k_z')]\right\}$$

$$= \frac{\sin(kR) - kR\cos(kR)}{k^3[k^2 - (\omega/c)^2]} \left\{\omega[\vec{\mathbb{P}}(\omega) \cdot \vec{\mathbb{E}}^{(\mathrm{i})}(k_z')][x(k_y \hat{\boldsymbol{z}} - k_z \hat{\boldsymbol{y}}) + y(k_z \hat{\boldsymbol{x}} - k_x \hat{\boldsymbol{z}}) + z(k_x \hat{\boldsymbol{y}} - k_y \hat{\boldsymbol{x}})]\right.$$
$$- \omega[\boldsymbol{k} \cdot \vec{\mathbb{E}}^{(\mathrm{i})}(k_z')]\{[x\mathbb{P}_y(\omega) - y\mathbb{P}_x(\omega)]\hat{\boldsymbol{z}} + z[\mathbb{P}_x(\omega)\hat{\boldsymbol{y}} - \mathbb{P}_y(\omega)\hat{\boldsymbol{x}}]\}$$
$$+ c\left\{(\omega/c)^2 [\vec{\mathbb{P}}(\omega) \cdot \vec{\mathbb{E}}^{(\mathrm{i})}(k_z')] - [\boldsymbol{k} \cdot \vec{\mathbb{P}}(\omega)][\boldsymbol{k} \cdot \vec{\mathbb{E}}^{(\mathrm{i})}(k_z')]\right\}(y\hat{\boldsymbol{x}} - x\hat{\boldsymbol{y}})$$
$$\left. + ck_z [\boldsymbol{k} \cdot \vec{\mathbb{P}}(\omega)]\left\{[x\mathbb{E}_y^{(\mathrm{i})}(k_z') - y\mathbb{E}_x^{(\mathrm{i})}(k_z')]\hat{\boldsymbol{z}} + z[\mathbb{E}_x^{(\mathrm{i})}(k_z')\hat{\boldsymbol{y}} - \mathbb{E}_y^{(\mathrm{i})}(k_z')\hat{\boldsymbol{x}}]\right\}\right\}$$

$$= \frac{\sin(kR) - kR\cos(kR)}{k^3[k^2 - (\omega/c)^2]} \left\{\omega[\mathbb{P}_x(\omega)\mathbb{E}_x^{(\mathrm{i})}(k_z') + \mathbb{P}_y(\omega)\mathbb{E}_y^{(\mathrm{i})}(k_z')][x(k_y \hat{\boldsymbol{z}} - k_z \hat{\boldsymbol{y}}) + y(k_z \hat{\boldsymbol{x}} - k_x \hat{\boldsymbol{z}}) + z(k_x \hat{\boldsymbol{y}} - k_y \hat{\boldsymbol{x}})]\right.$$
$$- \omega[k_x \mathbb{E}_x^{(\mathrm{i})}(k_z') + k_y \mathbb{E}_y^{(\mathrm{i})}(k_z')]\{[x\mathbb{P}_y(\omega) - y\mathbb{P}_x(\omega)]\hat{\boldsymbol{z}} + z[\mathbb{P}_x(\omega)\hat{\boldsymbol{y}} - \mathbb{P}_y(\omega)\hat{\boldsymbol{x}}]\}$$
$$+ c(\omega/c)^2 [\mathbb{P}_x(\omega)\mathbb{E}_x^{(\mathrm{i})}(k_z') + \mathbb{P}_y(\omega)\mathbb{E}_y^{(\mathrm{i})}(k_z')](y\hat{\boldsymbol{x}} - x\hat{\boldsymbol{y}})$$
$$- c[k_x^2 \mathbb{P}_x(\omega)\mathbb{E}_x^{(\mathrm{i})}(k_z') + k_y^2 \mathbb{P}_y(\omega)\mathbb{E}_y^{(\mathrm{i})}(k_z') + k_x k_y \mathbb{P}_x(\omega)\mathbb{E}_y^{(\mathrm{i})}(k_z') + k_y k_x \mathbb{P}_y(\omega)\mathbb{E}_x^{(\mathrm{i})}(k_z')](y\hat{\boldsymbol{x}} - x\hat{\boldsymbol{y}})$$
$$\left. + ck_z [k_x \mathbb{P}_x(\omega) + k_y \mathbb{P}_y(\omega)]\left\{[x\mathbb{E}_y^{(\mathrm{i})}(k_z') - y\mathbb{E}_x^{(\mathrm{i})}(k_z')]\hat{\boldsymbol{z}} + z[\mathbb{E}_x^{(\mathrm{i})}(k_z')\hat{\boldsymbol{y}} - \mathbb{E}_y^{(\mathrm{i})}(k_z')\hat{\boldsymbol{x}}]\right\}\right\}. \qquad \text{(E16)}$$

The integral $\iiint x \exp(\mathrm{i}k_x x) \exp(\mathrm{i}k_y y) \exp[\mathrm{i}(k_z + k_z')z] \, \mathrm{d}x\mathrm{d}y\mathrm{d}z$ yields the triplet of $\delta$-functions $-\mathrm{i}(2\pi)^3 \delta'(k_x)\delta(k_y)\delta(k_z' + k_z)$, which we shall, for the sake of simplicity, write as $-\mathrm{i}(2\pi)^3 \delta_x' \delta_y \delta_z$. Similar results are obtained when the integrand's leading $x$ is replaced by $y$ or $z$. We then have

$$\boldsymbol{\mathcal{L}}_{\text{cross}}(t_0)=\frac{-\mathrm{i}(2\pi)^3}{(2\pi)(4\pi^3)c^2}\int_{-\infty}^{\infty}\frac{\sin(kR)-kR\cos(kR)}{k^3[k^2-(\omega/c)^2]}\{\omega[\mathbb{P}_x(\omega)\mathbb{E}_x^{(\mathrm{i})}(k_z')+\mathbb{P}_y(\omega)\mathbb{E}_y^{(\mathrm{i})}(k_z')]$$

$$\times[(k_y\hat{\boldsymbol{z}}-k_z\hat{\boldsymbol{y}})\delta_x'\delta_y\delta_z+(k_z\hat{\boldsymbol{x}}-k_x\hat{\boldsymbol{z}})\delta_x\delta_y'\delta_z+(k_x\hat{\boldsymbol{y}}-k_y\hat{\boldsymbol{x}})\delta_x\delta_y\delta_z']$$

$$-\omega[k_x\mathbb{E}_x^{(\mathrm{i})}(k_z')+k_y\mathbb{E}_y^{(\mathrm{i})}(k_z')]\{[\mathbb{P}_y(\omega)\delta_x'\delta_y\delta_z-\mathbb{P}_x(\omega)\delta_x\delta_y'\delta_z]\hat{\boldsymbol{z}}+[\mathbb{P}_x(\omega)\hat{\boldsymbol{y}}-\mathbb{P}_y(\omega)\hat{\boldsymbol{x}}]\delta_x\delta_y\delta_z'\}$$

$$+c(\omega/c)^2[\mathbb{P}_x(\omega)\mathbb{E}_x^{(\mathrm{i})}(k_z')+\mathbb{P}_y(\omega)\mathbb{E}_y^{(\mathrm{i})}(k_z')](\delta_x\delta_y'\delta_z\hat{\boldsymbol{x}}-\delta_x'\delta_y\delta_z\hat{\boldsymbol{y}})$$

$$-c[k_x^2\mathbb{P}_x(\omega)\mathbb{E}_x^{(\mathrm{i})}(k_z')+k_y^2\mathbb{P}_y(\omega)\mathbb{E}_y^{(\mathrm{i})}(k_z')+k_xk_y\mathbb{P}_x(\omega)\mathbb{E}_y^{(\mathrm{i})}(k_z')+k_yk_x\mathbb{P}_y(\omega)\mathbb{E}_x^{(\mathrm{i})}(k_z')](\delta_x\delta_y'\delta_z\hat{\boldsymbol{x}}-\delta_x'\delta_y\delta_z\hat{\boldsymbol{y}})$$

$$+ck_z[k_x\mathbb{P}_x(\omega)+k_y\mathbb{P}_y(\omega)]\{[\mathbb{E}_y^{(\mathrm{i})}(k_z')\delta_x'\delta_y\delta_z-\mathbb{E}_x^{(\mathrm{i})}(k_z')\delta_x\delta_y'\delta_z]\hat{\boldsymbol{z}}+[\mathbb{E}_x^{(\mathrm{i})}(k_z')\hat{\boldsymbol{y}}-\mathbb{E}_y^{(\mathrm{i})}(k_z')\hat{\boldsymbol{x}}]\delta_x\delta_y\delta_z'\}\}$$

$$\times\exp(-\mathrm{i}ck_z't_0)\exp(-\mathrm{i}\omega t_0)\,\mathrm{d}k_z'\mathrm{d}\boldsymbol{k}\mathrm{d}\omega. \tag{E17}$$

**Digression**: The function $f(k)=[\sin(kR)-kR\cos(kR)]/\{k^3[k^2-(\omega/c)^2]\}$, evaluated at $k_x=k_y=0$, becomes $f(k_z)$. Also, $\partial_{k_x}f(k)=\partial_k f(k)\times\partial_{k_x}k=(k_x/k)\partial_k f(k)=0$, when evaluated at $k_x=0$. Similarly, $\partial_{k_y}f(k)=0$ at $k_y=0$. As for $\partial_{k_z'}\exp(-\mathrm{i}ck_z't_0)\big|_{k_z'=-k_z}=-\mathrm{i}ct_0\exp(\mathrm{i}ck_zt_0)$, the relevant coefficient turns out to be zero, as shown below.

$$\omega[\mathbb{P}_x(\omega)\mathbb{E}_x^{(\mathrm{i})*}(k_z)+\mathbb{P}_y(\omega)\mathbb{E}_y^{(\mathrm{i})*}(k_z)](k_x\hat{\boldsymbol{y}}-k_y\hat{\boldsymbol{x}})\delta(k_x)\delta(k_y)$$

$$-\omega[k_x\mathbb{E}_x^{(\mathrm{i})*}(k_z)+k_y\mathbb{E}_y^{(\mathrm{i})*}(k_z)][\mathbb{P}_x(\omega)\hat{\boldsymbol{y}}-\mathbb{P}_y(\omega)\hat{\boldsymbol{x}}]\delta(k_x)\delta(k_y)$$

$$+ck_z[k_x\mathbb{P}_x(\omega)+k_y\mathbb{P}_y(\omega)][\mathbb{E}_x^{(\mathrm{i})*}(k_z)\hat{\boldsymbol{y}}-\mathbb{E}_y^{(\mathrm{i})*}(k_z)\hat{\boldsymbol{x}}]\delta(k_x)\delta(k_y)\to 0.$$

$$\boldsymbol{\mathcal{L}}_{\text{cross}}(t_0)=\frac{\mathrm{i}}{\pi c^2}\int_{-\infty}^{\infty}\frac{\sin(k_zR)-k_zR\cos(k_zR)}{k_z^3[k_z^2-(\omega/c)^2]}\{[\omega\mathbb{P}_x(\omega)\mathbb{E}_y^{(\mathrm{i})*}(k_z)-\omega\mathbb{P}_y(\omega)\mathbb{E}_x^{(\mathrm{i})*}(k_z)]\hat{\boldsymbol{z}}$$

$$+ck_z[\mathbb{P}_x(\omega)\mathbb{E}_y^{(\mathrm{i})*}(k_z)-\mathbb{P}_y(\omega)\mathbb{E}_x^{(\mathrm{i})*}(k_z)]\hat{\boldsymbol{z}}\}\exp(\mathrm{i}ck_zt_0)\exp(-\mathrm{i}\omega t_0)\,\mathrm{d}k_z\mathrm{d}\omega$$

$$=\frac{\mathrm{i}}{\pi c^2}\int_{-\infty}^{\infty}\frac{\sin(k_zR)-k_zR\cos(k_zR)}{k_z^3[k_z^2-(\omega/c)^2]}[\omega\vec{\mathbb{P}}(\omega)+ck_z\vec{\mathbb{P}}(\omega)]\times\vec{\mathbb{E}}^{(\mathrm{i})*}(k_z)\exp(\mathrm{i}ck_zt_0)\exp(-\mathrm{i}\omega t_0)\,\mathrm{d}\omega\mathrm{d}k_z$$

$$=\frac{\mathrm{i}}{\pi c}\int_{-\infty}^{\infty}\frac{\sin(k_zR)-k_zR\cos(k_zR)}{k_z^4}\{2\pi\mathrm{i}[ck_z\vec{\mathbb{P}}_R(ck_z)\cos(ck_zt_0)+ck_z\vec{\mathbb{P}}_I(ck_z)\sin(ck_zt_0)]$$

$$+2\pi ck_z[\vec{\mathbb{P}}_R(ck_z)\sin(ck_zt_0)-\vec{\mathbb{P}}_I(ck_z)\cos(ck_zt_0)]\}\times\vec{\mathbb{E}}^{(\mathrm{i})*}(k_z)\exp(\mathrm{i}ck_zt_0)\,\mathrm{d}k_z$$

$$=2\mathrm{i}\int_{-\infty}^{\infty}\frac{\sin(k_zR)-k_zR\cos(k_zR)}{k_z^3}[\mathrm{i}\vec{\mathbb{P}}_R(ck_z)-\vec{\mathbb{P}}_I(ck_z)]\times\vec{\mathbb{E}}^{(\mathrm{i})*}(k_z)\mathrm{d}k_z$$

$$=-2\int_{-\infty}^{\infty}\frac{\sin(k_zR)-k_zR\cos(k_zR)}{k_z^3}\vec{\mathbb{P}}(ck_z)\times\vec{\mathbb{E}}^{(\mathrm{i})*}(k_z)\mathrm{d}k_z$$

$$=-2(\varepsilon_0/c)\int_{-\infty}^{\infty}\frac{\sin(k_zR)-k_zR\cos(k_zR)}{k_z^3}\chi(ck_z)\vec{\mathbb{E}}^{(\mathrm{i})}(k_z)\times\vec{\mathbb{E}}^{(\mathrm{i})*}(k_z)\mathrm{d}k_z. \tag{E18}$$

The above angular momentum is deducted from that of the incident packet, which remains intact as it propagates past the particle. If a fraction of the energy (and, therefore, angular momentum) of the incident beam is permanently absorbed within the particle, then the overall angular momentum content of the EM field will be reduced, while the particle picks up the deficit in the form of mechanical angular momentum. If, however, the particle happens to be transparent (i.e., $\gamma=0$), no net torque will be exerted on the particle, in which case the EM angular momentum contained in the dipole radiation, as given by Eq.(E12), will be precisely cancelled out by the cross-terms given by Eq.(E18).